\documentclass[journal,twoside]{IEEEtran}
\usepackage[explicit]{titlesec}
\usepackage{amsthm}
\usepackage{amsmath,amsfonts}
\usepackage{algorithmic}
\usepackage{algorithm}
\usepackage{array}
\usepackage{booktabs}
\usepackage{bm}
\usepackage[caption=false,font=footnotesize,labelfont=rm,textfont=rm]{subfig}
\usepackage{color}
\usepackage{mathtools}
\usepackage{setspace}
\usepackage{textcomp}
\usepackage{multirow}
\usepackage{threeparttable}
\usepackage{makecell}
\usepackage{stfloats}
\usepackage{url}
\usepackage{verbatim}
\usepackage{graphicx}
\usepackage{cite}
\usepackage{xcolor}
\usepackage{etoolbox}

\makeatletter
\newcommand{\normalnumbering}{%
  \renewcommand{\tagform@}[1]{(\ignorespaces##1\unskip)}%
}

\newcommand{\rednumbering}{%
  \renewcommand{\tagform@}[1]{\textcolor{red}{(\ignorespaces##1\unskip)}}%
}
\makeatother

\begin{document}
\title{Thévenin Equivalent Parameters Identification Based on Statistical Characteristics of System Ambient Data}

\author{
	Boying Zhou,~\IEEEmembership{Student Member,~IEEE}, Chen Shen,~\IEEEmembership{Senior Member,~IEEE}, and Kexuan Tang,~\IEEEmembership{Student Member,~IEEE}
	\thanks{This work was supported by the National Natural Science Foundation of China under Grant U23B6008. \textit{(Corresponding author: Chen Shen.)}}
	\thanks{The authors are with the Department of Electrical Engineering, Tsinghua University, Beijing 100084, China (e-mail: zby21@mails.tsinghua.edu.cn; shenchen@mail.tsinghua.edu.cn; 14z32@163.com).}
}

\markboth{IEEE transactions on power systems,~Vol.~XX, No.~X, XXX~XXXX}
{ZHOU \MakeLowercase{\textit{et al.}}: Thévenin Equivalent Parameters Identification Based on statistical characteristics of System Ambient Data}


\maketitle
\begin{abstract}
This paper proposes a novel method for identifying Thévenin equivalent parameters (TEP) in power system, based on the statistical characteristics of the system's stochastic response. The method leverages stochastic fluctuation data under steady-state grid conditions and applies sliding window techniques to compute sensitivity parameters between voltage magnitude, current magnitude and power. This enables high-accuracy and robust TEP identification. In contrast to traditional methods, the proposed approach does not rely on large disturbances or probing signals but instead utilizes the natural fluctuation behavior of the system. Additionally, the method supports distributed implementation using local measurements of voltage magnitude, current magnitude, and power, offering significant practical value for engineering applications. The theoretical analysis demonstrates the method's robustness in the presence of low signal-to-noise ratio (SNR), asynchronous measurements, and data collinearity issues. Simulation results further confirm the effectiveness of the proposed method in diverse practical scenarios, demonstrating its ability to consistently provide accurate and reliable identification of TEP using system ambient data.
\end{abstract}


\section{Introduction}

\subsection{Motivation}
\IEEEPARstart{T}{hévenin} equivalent (TE) serves as a core tool for assessing power system stability and safety. By simplifying complex networks into an equivalent voltage source and impedance, TE provides precise and efficient tools for steady-state analyses and transient fault diagnosis. With the widespread adoption of Phasor Measurement Unit (PMU) technology,  interest in online identification and monitoring of Thévenin equivalent parameters (TEP) using PMU data has increased \cite{Ref1_vu1999use}.

To this end, numerous PMU-based methods have been proposed to estimate TEP using routine measurement data. For example, existing techniques such as least-squares fitting~\cite{Ref7_burchett2018optimal}, robust regression~\cite{Ref9_su2018robust}, recursive constrained estimation~\cite{Ref12_fusco2000constrained}, and dynamic impedance modeling~\cite{Ref13_malkhandi2022dynamic} all exploit time-series relationships among voltage, current, and power. These methods assume that the system experiences sufficient natural variation over short time windows, allowing model parameters to be fitted from instantaneous input–output relationships.

While these methods do not require large disturbances or probing signals, they still rely on sufficient natural variability in the data. When the system operates in a highly stable regime—such as with near-constant load or voltage—the inherent signal fluctuations may be too small or strongly collinear, leading to ill-conditioned regression matrices with large condition numbers~\cite{Ref7_burchett2018optimal}. Such lack of excitation causes the information matrix to degrade, making the parameter estimation highly sensitive to noise and numerical error~\cite{Rawlings2016}. This structural instability is further exacerbated by a low signal-to-noise ratio (SNR). Under quasi-steady-state conditions, signal fluctuations may be comparable to or even weaker than background noise~\cite{Ref3_na2023gaussian}, which amplifies estimation variance and undermines the identifiability of TEP.

To address these limitations, this paper proposes a novel framework that identifies TEP based on the statistical characteristics of ambient fluctuations. Instead of relying on pointwise input-output dynamics, it treats voltage, current, and power variations as realizations of stationary stochastic processes~\cite{Ref2_milano2013systematic}. By applying a sliding window to the time series, key statistics are extracted as sensitivity indicators. These characteristics enable consistent parameter estimation even under low SNR, asynchronous data, or strong power correlation. This statistical approach ensures robust and accurate online estimation without requiring large signal variations, making it well suited for steady-state grid monitoring.

\subsection{Literature Review}

The method using a limited number of measurement samples is one of the simplest approaches for TEP identification, relying on local data from adjacent time points to identify model parameters \cite{Ref1_vu1999use}. While widely used for its simplicity, it assumes constant system parameters \cite{Ref4_weckx2015voltage}, limiting its ability to capture dynamic system behavior. It is also sensitive to the choice of measurement points, nonlinearity, and noise, which can lead to significant errors \cite{Ref5_kamel2021generalized}.

To overcome these limitations, several improved algorithms based on least squares (LS) have been proposed, including classical LS \cite{Ref7_burchett2018optimal}, \cite{Ref8_haji2018online}, robust LS \cite{Ref9_su2018robust},\cite{Ref10_zhou2008electromechanical}, recursive LS \cite{Ref11_wang2012real}, and constrained LS \cite{Ref12_fusco2000constrained}. These methods typically require large data windows, which can delay estimation. To address this, techniques such as the Kalman filter for dynamic updating \cite{Ref13_malkhandi2022dynamic}, extended Kalman filters \cite{Ref15_wang2018multi} and multivariate Huber loss function \cite{Ref16_peker2016fitting} have been used to reduce errors and improve robustness, especially under non-Gaussian noise.

In recent years, sensitivity analysis-based algorithms have emerged as a prominent approach for TEP identification. These methods estimate parameters based on the sensitivity relationship between electrical quantity fluctuations, making them well-suited for scenarios with frequent system dynamic changes \cite{Ref17_alinezhad2023online}. For example, Ref.\cite{Ref18_matavalam2018sensitivity} introduces the Sensitivity-Based Thévenin Index (STI) to monitor voltage stability, while Ref.\cite{Ref19_smon2006local} refines the STI using Tellegen's theorem. Ref.\cite{Ref20_zhang2018sensitivity} extends STI to assess voltage stability during N-1 transmission line faults.

However, nonlinearities, collinearity, and periodic characteristics of power systems often lead to insufficient data for accurate sensitivity identification \cite{Ref21_zhang2017power}. To address this, Ref.\cite{Ref22_chen2016measurement} applies Total Least Squares (TLS) to handle collinearity, Ref.\cite{Ref23_silva2020data} uses ridge regression with directional forgetting for stability during low-excitation, and Ref.\cite{Ref24_tang2024adaptive} combines local weighted ridge regression with regularization for stabilization. Ref.\cite{Ref25_liang2023temporally} uses adaptive weighted sparsity and Huber loss to reduce collinearity and handle non-Gaussian noise. Ref.\cite{Ref26_chang2022data} develops a nonlinear regression model for parameter interdependencies.

Despite advancements in addressing noise, collinearity, and data issues, these methods still face challenges under steady-state conditions, often resulting in significant errors, particularly in low SNR or minimal disturbance scenarios. Achieving efficient and accurate TEP identification using only stochastic fluctuation data remains a critical and unresolved challenge.


\subsection{Contribution}

This paper presents a new framework for identifying TEP based on the statistical characteristics of stochastic fluctuations observed under ambient operating conditions.   The proposed method models voltage, current, and power variations as realizations of stationary stochastic processes, and extracts their statistical characteristics within a sliding-window structure.  

The key contributions are as follows:

\begin{enumerate}
	\item \textbf{Statistical formulation for MSP identification under ambient conditions.} We propose a statistical approach for identifying MSP from random fluctuations in ambient data. Two complementary methods are developed within a sliding-window framework: one based on windowed means~\eqref{Eqn9}, and the other on variances and covariances~\eqref{Eqn10}. These MSPs quantify the sensitivity  serve as intermediate indicators for subsequent TEP estimation (Section~II-A).

	\item \textbf{Theoretical errors of the proposed method in various scenarios.} We derive error expressions under challenging conditions such as low SNR, asynchronous measurements, and high data collinearity. To quantify performance, key indicators including the signal-to-noise ratio (SNR), variance-based deviation ratio, and condition number of the processed data are analyzed (Section~II-B).

	\item \textbf{Model selection strategy for low-quality data scenarios.} Based on the derived error bounds and condition number criteria, we compare two proposed identification methods and offer a principled selection strategy for choosing the one with the best expected performance under a given data quality condition (e.g., low SNR or high collinearity)(Section~III-C and Fig.~\ref{Fig_err}).

    \item \textbf{Comprehensive robustness evaluation under realistic conditions.} The proposed method is validated using enhanced versions of the IEEE 39-bus system and the CSEE-RAS system  (Section~IV-B), including the integration of renewable energy sources and dynamic loads to assess robustness. In addition, we evaluate its applicability under a variety of practical scenarios, including non-Gaussian measurement noise, network reconfiguration, bad data injection, and missing data (Section~V).

\end{enumerate}

The remainder of this paper is organized as follows.
Section II introduces the proposed TEP identification framework based on MSP. Section III details the statistical formulation and analyzes its theoretical performance under low SNR, asynchronous sampling, and collinearity. Section IV presents simulation-based validations. Section V concludes the paper.

\section{TEP Identification Based on Magnitude Sensitivity}

\subsection{Thévenin Equivalent Model}
\begin{figure}[!htbp] 
	\centering	\includegraphics[width=0.8\columnwidth]{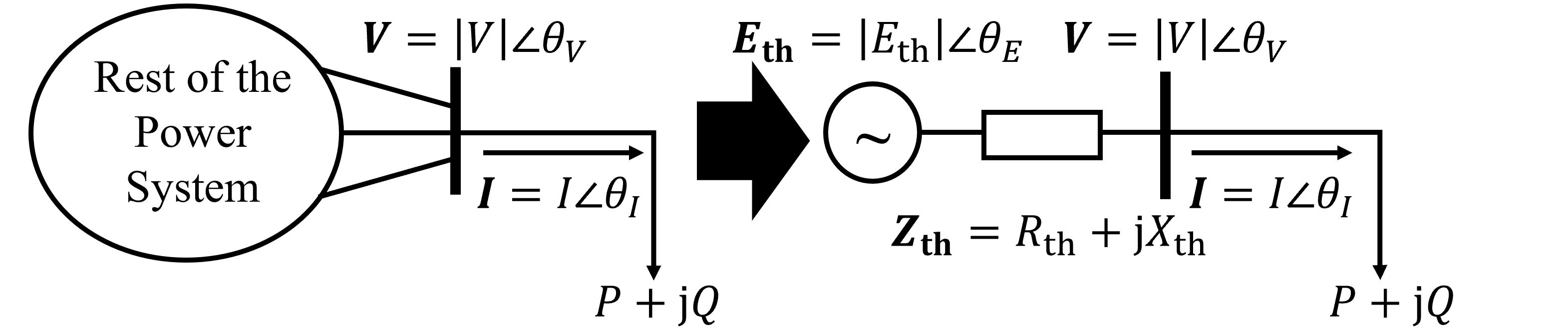}
	\caption{Schematic diagram of Thévenin equivalent in a power system.}
	\label{Fig1_Thevenin_equivalent}
\end{figure}

The schematic diagram for Thevenin equivalent (TE) in a power system is shown in Figure 1. This method divides the power system at the equivalent port (load node or flow interface) into two parts: the ‘equivalent side’ and the ‘load side’. The ‘equivalent side’ is represented as a TE voltage source in series with a TE impedance, as shown in Fig.~\ref{Fig1_Thevenin_equivalent}.

In Fig.~\ref{Fig1_Thevenin_equivalent}, the TEP used for voltage stability analysis are ${\left|E_\text{th}\right|}$, $R_{\text{th}}$, and $X_{\text{th}}$. According to Kirchhoff's Voltage Law and the power equations of the equivalent circuit \cite{kundur1994power}, the phasor expressions at the equivalent port are given by
\begin{equation} \label{Eqn-add-1}
	\begin{cases}
		\bm E_\text{th}=\bm I\cdot \bm Z_\text{th}+\bm V \\
		P+\text{j}Q=\bm V\cdot\bm I^*,
	\end{cases}
\end{equation}
which can be transformed into magnitude-domain form as
\begin{equation} \label{Eqn-add-2}
\resizebox{1.0\columnwidth}{!}{$
	\begin{cases}
        \begin{aligned}
            {{\left| V \right|}^{4}} &+\left( R_{\text{th}}^{2}+X_{\text{th}}^{2} \right)\left( {{P}^{2}}+{{Q}^{2}} \right)
            
            +\left( 2{{R}_{\text{th}}}P+2{{X}_{\text{th}}}Q-{{\left| {{E}_{\text{th}}} \right|}^{2}} \right){{\left| V \right|}^{2}}=0
        \\
             \big( R_{\text{th}}^{2}&+X_{\text{th}}^{2}\big)\,{{\left| I \right|}^{4}}+\left(P^2 + Q^2\right)
            +\left( 2{{R}_{\text{th}}}P+2{{X}_{\text{th}}}Q-{{\left| {{E}_{\text{th}}} \right|}^{2}} \right){{\left| I \right|}^{2}}=0.
        \end{aligned}
    \end{cases}
    $}
\end{equation}
Each equation in \eqref{Eqn-add-2} is quadratic in $\left| V \right|^2$ and $\left| I \right|^2$, respectively. By solving them analytically, the closed-form solutions can be obtained as
\begin{equation} \label{Eqn1}
	\begin{cases}
		\left| V \right|=\sqrt{\frac{{{\left| {{E}_{\text{th}}} \right|}^{2}}-2({{R}_{\text{th}}}P+{{X}_{\text{th}}}Q)+\sqrt{\Delta }}{2}} \\
		\left| I \right|=\sqrt{\frac{{{\left| {{E}_{\text{th}}} \right|}^{2}}-2(P{{R}_{\text{th}}}+Q{{X}_{\text{th}}})-\sqrt{\Delta }}{2(R_{\text{th}}^{2}+X_{\text{th}}^{2})}} \\ 
	\end{cases}
\end{equation}
where discriminant $\Delta ={{\left| {{E}_{\text{th}}} \right|}^{4}}-4\left( {{R}_{\text{th}}}P+{{X}_{\text{th}}}Q \right){{\left| {{E}_{\text{th}}} \right|}^{2}}-4{{\left( {{X}_{\text{th}}}P-{{R}_{\text{th}}}Q \right)}^{2}}$. 

\subsection{MSP Based Identification of TEP}
Considering that sensors at the boundaries of distribution and transmission networks typically cannot measure voltage phase angles \cite{Ref27_talkington2024conditions}, this paper uses magnitude sensitivity for TEP identification to enhance the method's applicability. Sensitivity analysis assumes that, within a small range, voltage and current magnitude changes exhibit a linear response to variations in active and reactive power, as shown as
\begin{equation} \label{Eqn2}
	\begin{bmatrix}
		{\Delta \left| V \right|} \\ {\Delta \left| I \right|} 
	\end{bmatrix} \!\approx\! \begin{bmatrix}
		\frac{\partial \left| V \right|}{\partial P} & \frac{\partial \left| V \right|}{\partial Q} \\
		\frac{\partial \left| I \right|}{\partial P} & \frac{\partial \left| I \right|}{\partial Q} 
	\end{bmatrix} \!\! \begin{bmatrix}
		{\Delta P} \\ {\Delta Q}
    \end{bmatrix} \!=\! \begin{bmatrix}
		{{\beta }_{\left| V \right|P}} & {{\beta }_{\left| V \right|Q}} \\
		{{\beta }_{\left| I \right|P}} & {{\beta }_{\left| I \right|Q}}
	\end{bmatrix} \!\! \begin{bmatrix}
		{\Delta P} \\ {\Delta Q}
	\end{bmatrix}
\end{equation}
where ${\beta }_{\left| V \right|P}$, ${\beta }_{\left| V \right|Q}$, ${\beta }_{\left| I \right|P}$ and ${\beta }_{\left| I \right|Q}$ are collectively called the magnitude sensitivity parameters (MSP). Since the partial derivatives matrix can be calculated from \eqref{Eqn1}, the theoretical MSP are given as
\begin{equation} \label{Eqn3}
\resizebox{1.0\columnwidth}{!}{$
	\begin{cases}
		{{\beta }_{\left| V \right|P}}=-\frac{\left( R_{\text{th}}^{2}+X_{\text{th}}^{2} \right){{{\tilde{P}}}_{k}}+{{R}_{\text{th}}}{{\left| {{{\tilde{V}}}_{k}} \right|}^{2}}}{\left| {{{\tilde{V}}}_{k}} \right|\sqrt{\tilde{\Delta}}} ,
        {{\beta }_{\left| I \right|P}}=\frac{{{{\tilde{P}}}_{k}}+{{R}_{\text{th}}}{{\left| {{{\tilde{I}}}_{k}} \right|}^{2}}}{\left| {{{\tilde{I}}}_{k}} \right|\sqrt{\tilde{\Delta}}} \\
        {{\beta }_{\left| V \right|Q}}=-\frac{\left( R_{\text{th}}^{2}+X_{\text{th}}^{2} \right){{{\tilde{Q}}}_{k}}+{{X}_{\text{th}}}{{\left| {{{\tilde{V}}}_{k}} \right|}^{2}}}{\left| {{{\tilde{V}}}_{k}} \right|\sqrt{\tilde{\Delta}}} ,        
        {{\beta }_{\left| I \right|Q}}=\frac{{{{\tilde{Q}}}_{k}}+{{X}_{\text{th}}}{{\left| {{{\tilde{I}}}_{k}} \right|}^{2}}}{\left| {{{\tilde{I}}}_{k}} \right|\sqrt{\tilde{\Delta}}}
	\end{cases}
    $}
\end{equation}
where discriminant $\tilde{\Delta} ={{\left| {{E}_{\text{th}}} \right|}^{4}}-4({{R}_{\text{th}}}{{\tilde{P}}_{k}}+{{X}_{\text{th}}}{{\tilde{Q}}_{k}}){{\left| {{E}_{\text{th}}} \right|}^{2}}-4{{({{X}_{\text{th}}}{{\tilde{P}}_{k}}-{{R}_{\text{th}}}{{\tilde{Q}}_{k}})}^{2}}$. ${\left| {{{\tilde{V}}}_{k}} \right|}$, ${\left| {{{\tilde{I}}}_{k}} \right|}$, ${{{\tilde{P}}}_{k}}$, and ${{{\tilde{Q}}}_{k}}$ represent the real-time measurements of electrical quantities. \eqref{Eqn3} reflects the nonlinear relationship between MSP and TEP.

When MSP are accurately identified, the true values of the TEP can be determined as the common solution of the theoretical equations \eqref{Eqn1} and MSP equations \eqref{Eqn3}. Based on the identification results of MSP and the real-time measurement data, the Levenberg-Marquardt method is applied for nonlinear least squares optimization to solve the system of equations \eqref{Eqn1} and \eqref{Eqn3}. The optimal solution yields the TEP values.

The proposed identification method calculates TEP in real-time using accurate MSP, based on Kirchhoff’s laws and power definitions, without requiring additional assumptions on load conditions. It is applicable to various power system equipment, offering higher accuracy than phase measurements and ensuring adaptability and versatility.

The accuracy of the method largely depends on the precision of MSP identification. The next chapter focuses on the core issue of accurately identifying MSP under the condition of stochastic fluctuations on the load side.

\section{Precise Sensitivity Identification Based on Stochastic Response statistical characteristics}

\begin{figure}[!htbp]
	\centering	\includegraphics[width=1.0\columnwidth]{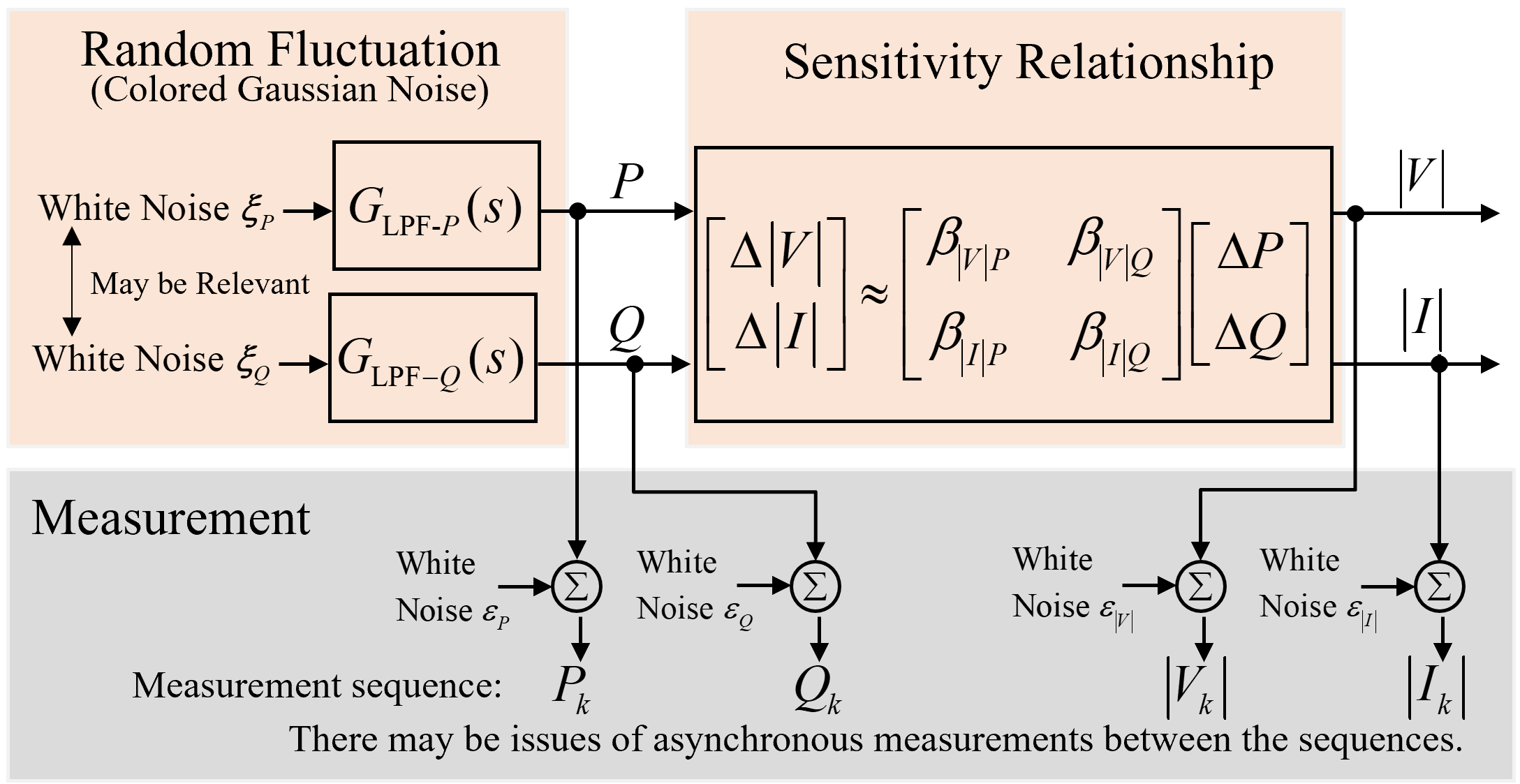}
	\caption{Port magnitude sensitivity relationship under steady-state conditions with stochastic load fluctuations.
    The random fluctuations of active and reactive power ($P$, $Q$) are modeled as colored Gaussian processes and propagate to voltage and current magnitudes ($|V|$, $|I|$) through a linear sensitivity relationship.
    All signals are further corrupted by Gaussian white measurement noise.
    The model also accounts for possible asynchronous sampling between different measurement sequences.
}
	\label{Fig2_Sensitivity}
\end{figure}

Under steady-state conditions in the power system, the random fluctuations of load power $P$ and $Q$ induce corresponding random fluctuations in voltage $\left|V\right|$ and current $\left|I\right|$, with their relationship described by \eqref{Eqn2}. The random fluctuations of $P$, $Q$, $\left|V\right|$, and $\left|I\right|$ are superimposed with measurement errors $\varepsilon _{P}$, $\varepsilon _{Q}$, $\varepsilon _{\left| V \right|}$, and $\varepsilon _{\left|I\right|}$, which follow Gaussian White Noise (GWM) characteristics. The random fluctuations of $P$ and $Q$ can be treated as a colored Gaussian process, characterized by autocorrelation between sample points, which can be modeled by passing GWN through a low-pass filter (such as the Ornstein-Uhlenbeck (O-U) process). Additionally, there may be cross-correlation between the random fluctuations of $P$ and $Q$. Under such steady-state conditions, the port's sensitivity relationship is shown in Fig.~\ref{Fig2_Sensitivity}.

MSP can be approximated using numerical differentiation methods, specifically through magnitude identification based on temporal increments \cite{Ref27_talkington2024conditions}, as shown as
\begin{equation} \label{Eqn4}
	\begin{bmatrix}
		\left| {{V}_{k+1}} \right|-\left| {{V}_{k}} \right|  \\
        \left| {{I}_{k+1}} \right|-\left| {{I}_{k}} \right| 
	\end{bmatrix} \!\approx\! \begin{bmatrix}
		{{\beta }_{\left| V \right|P}} & {{\beta }_{\left| V \right|Q}}  \\
        {{\beta }_{\left| I \right|P}} & {{\beta }_{\left| I \right|Q}}
	\end{bmatrix} \!\! \begin{bmatrix}
		{{P}_{k+1}}-{{P}_{k}}  \\
        {{Q}_{k+1}}-{{Q}_{k}}
	\end{bmatrix}
\end{equation}
where $k=1,2,\cdots ,n$ denotes the $k$-th sampled data point. The actual data sequence used for this identification is $Y_k=X_{k+1}-X_k$, $X=\left| V \right|,\left| I \right|,P,Q$. \eqref{Eqn4} allows using methods like LS to approximate MSP. Measurement errors and noise significantly affect estimation, especially when only ambient data is available. Therefore, enhancements to the method are necessary to mitigate low SNR effects, reduce parameter bias, and improve robustness against stochastic disturbances.




\subsection{Identification Based on statistical characteristics}
\subsubsection{statistical characteristics under Sliding Window}
To ensure that the extracted statistical characteristics are both meaningful and robust, we adopt a sliding window strategy. Since the proposed method relies on the statistical properties (mean, variance, covariance) of ambient electrical fluctuations, it is essential to compute these features over local data segments rather than over the entire time series. The sliding window approach enables localized estimation that balances noise suppression with responsiveness to slowly changing system conditions.

\begin{figure}[!htbp]
	\centering	
    \includegraphics[width=0.85\columnwidth]{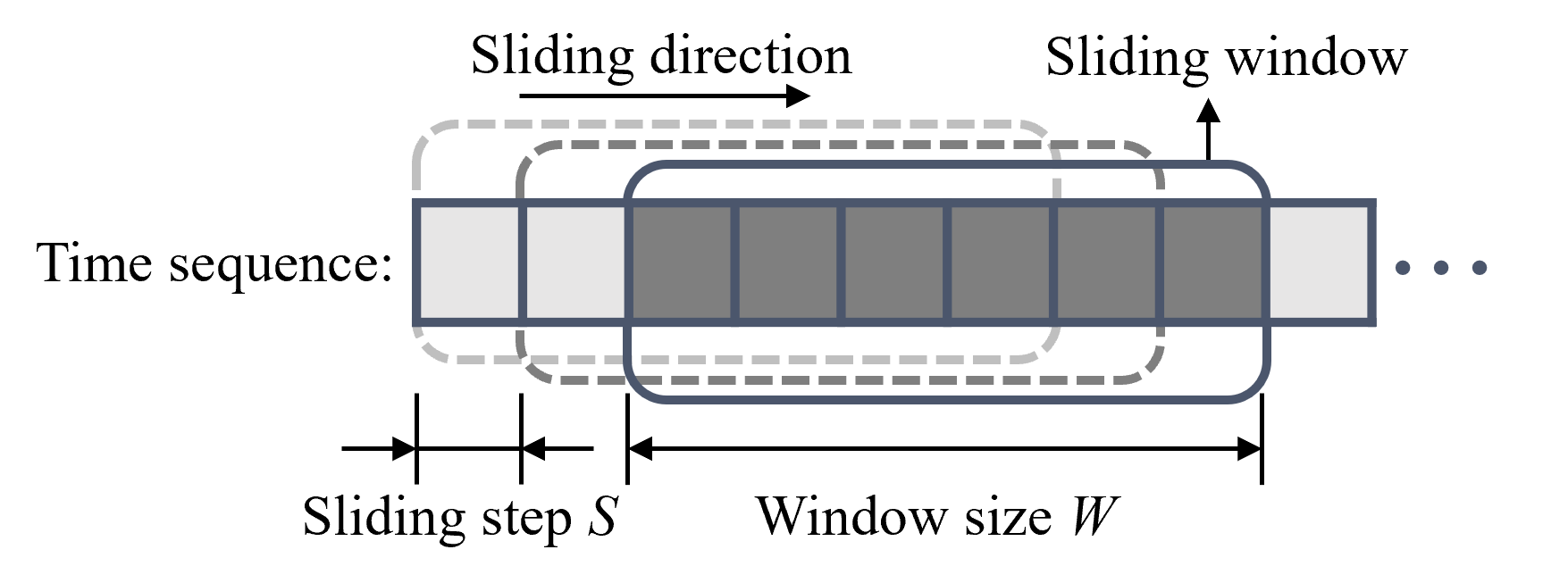}
	\caption{Sliding window diagram. } 
	\label{Fig3_Sliding_Window}
\end{figure}

To extract the statistical characteristics of the time sequence of electrical quantities, the sliding window technique is applied to process the stochastic fluctuation data. By defining a sliding window, the time series is divided into smaller sub-windows, as shown in Fig.~\ref{Fig3_Sliding_Window}.

Given the sampling period of the port electrical quantities as $T_{\text{s}}$, and to improve data utilization, the sliding step size is set as $S=T_{\text{s}}$. With the window size $W$, the number of data points within each sliding window is $n=WT_{\text{s}}^{-1}$, allowing for the acquisition of $n$ measurements of the system's electrical quantities, denoted as $\left[ {{X}_{k}},{{X}_{k+1}},\cdots ,{{X}_{k+n-1}} \right]$, $X=\left| V \right|,\left| I \right|,P,Q$. This approach replaces the original data with the statistical characteristics of the data within the sliding window for identification. The statistical characteristics data used for identification in the $k$-th sliding window can be expressed as
\begin{equation} \label{Eqn3a} 
{Y_k}=f\left( {X_k},{X_{k+1}},\cdots ,{X_{k+n-1}} \right),\,X=\left| V \right|,\left| I \right|,P,Q
\end{equation} 
where $f$ is the statistical function.
When $f$ is the mean function, the mean of data in the $k$-th window is
\begin{equation}\label{Eqn3b}  
    {{\bar{X}}_{\text{w,}k}}=\frac{1}{n}\sum\limits_{i=k}^{k+n-1}{{{X}_{i}}},\,X=\left| V \right|,\left| I \right|,P,Q.
\end{equation} 
When $f$ is the variance function, the variance of data in the $k$-th window is
\begin{equation} \label{Eqn3c} 
    s_{X,k}^{2}=\frac{1}{n-1}\sum\limits_{i=k}^{k+n-1}{{{\left( {{X}_{i}}-{{{\bar{X}}}_{\text{w,}k}} \right)}^{2}}},\,X=\left| V \right|,\left| I \right|,P,Q.
\end{equation} 
When $f$ is the covariance function, the covariance of $P$ and $Q$ in the $k$-th window is 
\begin{equation} \label{Eqn3d}
    {{s}_{PQ,k}}=\frac{1}{n-1}\sum\limits_{i=k}^{k+n-1}{\left[ \left( {{P}_{i}}-{{{\bar{P}}}_{\text{w,}k}} \right)\left( {{Q}_{i}}-{{{\bar{Q}}}_{\text{w,}k}} \right) \right]}.
\end{equation} 
\subsubsection{Mean-Based Identification under Sliding Window}
From the time-incremental identification method shown in \eqref{Eqn4}, we can further derive the sensitivity relationship based on the statistical characteristics of the temporal fluctuations.

First, the sensitivity relationship based on the mean within the sliding window can be derived. By summing the time-incremental changes within the window, the same linear sensitivity relationship is maintained as
\begin{equation} \label{Eqn9}
	\begin{bmatrix}
		\left| {{V}_{k}} \right|-\left| {{{\bar{V}}}_{\text{w,}k}} \right|  \\
        \left| {{I}_{k}} \right|-\left| {{{\bar{I}}}_{\text{w,}k}} \right|   
	\end{bmatrix} \!\approx\! \begin{bmatrix}
		{{\beta }_{\left| V \right|P}} & {{\beta }_{\left| V \right|Q}} \\
        {{\beta }_{\left| I \right|P}} & {{\beta }_{\left| I \right|Q}}
	\end{bmatrix} \!\! \begin{bmatrix}
		{{P}_{k}}-{{{\bar{P}}}_{\text{w,}k}}  \\
        {{Q}_{k}}-{{{\bar{Q}}}_{\text{w,}k}}  
	\end{bmatrix}.
\end{equation}
As shown in \eqref{Eqn9}, the mean and increment relative to the mean, are computed from original data. These feature data follow the same MSP identification relationship as in \eqref{Eqn4}. The data in the $k$-th sliding window is ${Y_k}={X_k}-{\bar{X}_{\text{w,}k}}$, $X=\left|V\right|,\left|I\right|,P,Q$.

\subsubsection{Variance-Based Identification under Sliding Window}
Furthermore, the sensitivity relationship based on the variance within the sliding window can be derived. By squaring and summing the deviations of the instantaneous values from the mean within the window, as shown in \eqref{Eqn9}, a linear sensitivity relationship between the variances and covariances of the electrical quantity samples is obtained. The sensitivity identification method based on window variance is then formulated as follows.
\begin{equation} \label{Eqn10}
	\begin{bmatrix}
		s_{\left| V \right|,k}^{2}  \\
        s_{\left| I \right|,k}^{2}    
	\end{bmatrix} \!\approx\! \begin{bmatrix}
		\beta _{\left| V \right|P}^{2} & \beta _{\left| V \right|Q}^{2} & 2{{\beta }_{\left| V \right|P}}{{\beta }_{\left| V \right|Q}}  \\
        \beta _{\left| I \right|P}^{2} & \beta _{\left| I \right|Q}^{2} & 2{{\beta }_{\left| I \right|P}}{{\beta }_{\left| I \right|Q}}
	\end{bmatrix} \!\! \begin{bmatrix}
		s_{P,k}^{2}  \\
        s_{Q,k}^{2}  \\
        {{s}_{PQ,k}}    
	\end{bmatrix} \!+\! \begin{bmatrix}
		{{\mu }_{{{\epsilon }_{1}}}}  \\
        {{\mu }_{{{\epsilon }_{2}}}}   
    \end{bmatrix}\!
\end{equation}
In \eqref{Eqn10}, ${{\mu }_{{{\epsilon }_{1}}}}$ and $ {{\mu }_{{{\epsilon }_{2}}}}$ represent the expected values of the identification errors. When the fluctuations of active power and reactive power are cross-correlated, ${{s}_{PQ,k}}\ne0$. 
As shown in \eqref{Eqn10}, the variance and covariance are computed from the original data. These feature data follow a new identification relationship for MSP. The data used in the $k$-th sliding window ${Y_k}$ is variance data $s_{X,k}^{2}$, $X=\left|V\right|,\left|I\right|,P,Q$, and covariance data ${{s}_{PQ,k}}$.

\subsection{Comparison of Identification Methods Performance}

This section compares three methods for identifying magnitude sensitivity parameters (MSP):
\begin{itemize}
    \item the \textbf{Baseline method}, based on temporal increments as in \eqref{Eqn4};
    \item the \textbf{Proposed mean-based method (Prop.1)}, using sliding-window centered data as in \eqref{Eqn9};
    \item the \textbf{Proposed variance-based method (Prop.2)}, using statistical moments within the sliding window as in \eqref{Eqn10}.
\end{itemize}

All three formulations can be interpreted as instances of the linear identification model $\mathbf{A} \boldsymbol{\Theta} = \mathbf{B}$. For clarity, Table~I summarizes the corresponding identification formulations. Both $\mathbf{A}_k$ and $\mathbf{B}_k$ are constructed from measurement data: $\mathbf{A}_k$ contains power features ($P$, $Q$), while $\mathbf{B}_k$ contains voltage and current magnitude features ($|V|$, $|I|$). $\boldsymbol{\Theta}$ denotes the unknown sensitivity parameters to be estimated.

Theoretical analysis will assess their performance in terms of SNR improvements, robustness to asynchronous measurements, and handling of data collinearity.

\begin{table*}[!htbp]
\centering
\caption{{Formulation of Three Identification Methods Based on Measured Data}}
\label{Tab:Formulation Methods}
\begin{tabular}{@{}lccc@{}}
\toprule
\textbf{Method} &
  $\mathbf{A}_k$ (Power Features) &
  $\boldsymbol{\Theta}$ (Sensitivity Parameters) &
  $\mathbf{B}_k$ (Voltage/Current Features) \\
\midrule

\textbf{{Baseline}} (Temporal Increments) &
  $ \begin{bmatrix} P_{k+1} - P_k & Q_{k+1} - Q_k \end{bmatrix} $ &
  $ \begin{bmatrix}
     \beta_{|V|P} & \beta_{|I|P}  \\
     \beta_{|V|Q} & \beta_{|I|Q}
   \end{bmatrix} $ &
  $ \begin{bmatrix} |V_{k+1}| - |V_k| & |I_{k+1}| - |I_k| \end{bmatrix} $ \\ \addlinespace

\textbf{{Prop.1}} (Mean-Based) &
  $ \begin{bmatrix} P_k - \bar{P}_{w,k} & Q_k - \bar{Q}_{w,k} \end{bmatrix} $ &
  same as above &
  $ \begin{bmatrix} |V_k| - \bar{|V|}_{w,k} & |I_k| - \bar{|I|}_{w,k} \end{bmatrix} $ \\ \addlinespace

\textbf{{Prop.2}} (Variance-Based) &
  $ \begin{bmatrix} s^2_{P,k} & s^2_{Q,k} & s_{PQ,k} \end{bmatrix} $ &
  $ \begin{bmatrix}
    \beta_{|V|P}^2 & \beta_{|I|P}^2 \\
    \beta_{|V|Q}^2 & \beta_{|I|Q}^2 \\
    2\beta_{|V|P} \beta_{|V|Q} & 2\beta_{|I|P} \beta_{|I|Q}
  \end{bmatrix} $ &
  $ \begin{bmatrix} s^2_{|V|,k} & s^2_{|I|,k} \end{bmatrix} $ \\

\bottomrule
\end{tabular}
\end{table*}

\subsubsection{Improvement in SNR}
In power systems, stochastic disturbances in electrical quantities are usually modeled as a stationary Gaussian process, considering temporal variations and correlations. The autocovariance function is ${{R}_{X}}\left( \tau  \right)=\operatorname{cov}\left[ X\left( t \right),X\left( t+\tau  \right) \right]$ with $\tau$ representing the time difference. The autocovariance function normalized by variance ${{R}_{X}}\left( 0  \right)=\sigma _{X}^{2}$ is $\rho \left( \tau  \right) $. For low-pass Gaussian colored noise, $\rho\left(\tau\right)$ decays or oscillates with increasing $\tau$. The $\tau$ required for $\rho\left(\tau\right)$ to decay to ${\text{e}}^{-1}$ of its initial value is the autocovariance time constant ${\tau }_{c}$.

In a sliding window with uniform sampling at a fixed interval ${T}_{\text{s}}$, the $n$ stochastic samples at $n$ time steps form an $n$-dimensional stochastic variable $\bm{X}={{\left[ {{X}_{1}},{{X}_{2}},\cdots ,{{X}_{n}} \right]}^{\text{T}}}$ . When the system is stable, these variables exhibit the same statistical properties, with temporal correlations present due to the nature of electrical quantities in power systems.

The covariance matrix of $\bm{X}$ is denoted as $\mathbf{C_X}$. $\mathbf{C_X}$ is symmetric and Toeplitz. The relationship between the elements in $\mathbf{C_X}$ and the autocovariance function $\rho \left( \tau  \right) $ is 
\begin{equation*}
    {C_X}_{(i,j)} = \sigma_X^2 \cdot \rho\left( |i - j| T_s \right), \quad i, j = 1, 2, \dots, n
\end{equation*}

The measurement errors $\varepsilon \left( t \right)$ is zero-mean GWN  with variance $\sigma _{\varepsilon }^{2}$. After applying the sliding window, $\varepsilon \left( t \right)$ also becomes an $n$-dimensional stochastic variable, with independent components, each following a zero-mean Gaussian distribution. Thus the system initial SNR can be defined using variance as 
\begin{equation}\label{Eqn11}
    {{\text{SNR}}_\textup{raw}}=10{{\log }_{10}}\left[ \frac{\operatorname{var}\left( X \right)}{\operatorname{var}\left( \varepsilon  \right)} \right]=10{{\log }_{10}}\left[ \frac{\sigma _{X}^{2}}{\sigma _{\varepsilon }^{2}} \right].
\end{equation}

The following compares the SNR of the data used in the three identification methods. For the baseline method based on temporal increments, the theoretical variance of the temporal increments sequence ${{Y}_{k}}={{X}_{k+1}}-{{X}_{k}}$ is 
\begin{equation}\label{Eqn12}
    \operatorname{var}\left( {{X}_{k+1}}-{{X}_{k}} \right)=2\sigma _{X}^{2}\left( 1-\rho \left( {{T}_{\text{s}}} \right) \right).
\end{equation}
As for the measurement error, $\operatorname{var}\left( {{\varepsilon }_{k+1}}-{{\varepsilon }_{k}}\right)$ is $2\sigma _{\varepsilon }^{2}$. Therefore, when calculating the sensitivity based on temporal increments, the theoretical value of the SNR for the data is
\begin{equation}\label{Eqn13}
    \begin{aligned}
        \text{SN}{{\text{R}}_{0}} 
        &=10{{\log }_{10}}\left[ \frac{\operatorname{var}\left( {{X}_{k+1}}-{{X}_{k}} \right)}{\operatorname{var}\left( {{\varepsilon }_{k+1}}-{{\varepsilon }_{k}}  \right)} \right]
       \\&=\text{SN}{{\text{R}}_\textup{raw}}+10{{\log }_{10}}\left( 1-\rho \left( {{T}_{\text{s}}} \right) \right).
    \end{aligned}
\end{equation}
When ${T_{\text{s}}}\ll {\tau _c}$, $\rho \left( {{T}_{\text{s}}} \right)\approx 1$. The data's SNR significantly decreases, which severely affects the accuracy of the least squares identification results.

For the proposed methods based on statistical characteristics under the sliding window, the statistical characteristics are also stochastic variables. 
For simplicity and clarity, we denote the single summation $\sum_{i=1}^{n}$ as $\sum_i$, $\sum_{j=1}^{n}$ as $\sum_j$, and the double summation $\sum_{i=1}^{n} \sum_{j=1}^{n}$ as $\sum_{i,j}$ throughout the paper. The theoretical variance of data sequence used by the mean-based method (Prop.1) ${{Y}_{k}}={{X}_{k}}-{{\bar{X}}_{\text{w,}k}}$, is
\begin{equation}\label{Eqn14}
    \begin{aligned}
        \operatorname{var}\big( X_k - \bar{X}_{\textup{w},k}\big) = &C_{X(1,1)} 
        - \frac{2}{n} \sum\nolimits_j C_{X(1,j)}  
        \\&
        + \frac{1}{n^2} \sum\nolimits_{i,j} C_{X(i,j)}.  
    \end{aligned}
\end{equation}

When $W\gg {{\tau }_{c}}\gg {{T}_{\text{s}}}$, the variance can be quickly approximated as
\begin{equation}\label{Eqn15}
    \operatorname{var}\left( X_k - \bar{X}_{\textup{w},k} \right)\approx \sigma _{X}^{2}\left( 1-\frac{2}{n{{T}_{\text{s}}}}\int_{{nT_{\text{s}}}/2}^{n{{T}_{\text{s}}}}{\rho \left( \tau  \right)}\text{d}\tau  \right)\approx \sigma _{X}^{2}.
\end{equation}
The data sequence of measurement noise is ${{\varepsilon }_{k}}-{{\bar{\varepsilon }}_{\text{w},k}}$, and its theoretical variance can be considered as a special case of \eqref{Eqn14}, given as $\frac{n-1}{n}\sigma _{\varepsilon }^{2}$.
Therefore, the SNR corresponding to the theoretical variance of the data sequence obtained by Prop.1 is:
\begin{equation}\label{Eqnextra}
\text{SN}{{\text{R}}_{1}}=10{{\log }_{10}}\left[ \frac{\operatorname{var}\left( X_k - \bar{X}_{\textup{w},k} \right)}{\operatorname{var}\left( {{\varepsilon }_{k}}-{{\bar{\varepsilon }}_{\text{w},k}}  \right)} \right]
\approx \text{SN}{{\text{R}}_\textup{raw}}.
\end{equation}

The data sequence used by the variance-based method (Prop.2) is $s_{X,k}^{2}$ ($X=P,Q$) and $s_{PQ,k}$. Assuming that the measurement noises in $P$ and $Q$ are independent, thus the SNR related to the covariance $s_{PQ,k}$ is not a concern. The theoretical variance of $s_{X,k}^{2}$ is
\begin{equation}\label{Eqn16}
        \operatorname{var}\left( s_{X,k}^{2} \right) = \frac{2}{(n-1)^2} \left( 
        S_{1a} - \frac{2}{n} S_{2a} + \frac{1}{n^2} S_{3a} 
        \right) 
\end{equation}
where $S_{1a} = \sum_{i,j}{C_{X\left( i,j \right)}^{2}}$, $S_{2a} = \sum_{i}{{{\left( \sum_{j}{{{C}_{X\left( i,j \right)}}} \right)}^{2}}}$, {$S_{3a} ={{\left( \sum_{i,j} {{{C}_{X\left( i,j \right)}}} \right)}^{2}}$. When $W\gg {{\tau }_{c}}\gg {{T}_{\text{s}}}$, the variance of  $s_{X,k}^{2}$ can be quickly approximated as
\begin{equation}\label{Eqn17}
    \operatorname{var}\left( s_{X,k}^{2} \right)\approx \frac{2\sigma _{X}^{4}}{\left( n-1 \right){{T}_{\text{s}}}}\int_{-{n{{T}_{\text{s}}}}/2}^{{n{{T}_{\text{s}}}}/2}{\rho {{\left( \tau  \right)}^{2}}}\text{d}\tau.
\end{equation}
The data sequence of measurement noise $s_{\varepsilon,k}^{2}$ follows a chi-square distribution. Its theoretical variance can be considered as a special case of \eqref{Eqn16}, given as $\frac{2}{n-1}\sigma _{\varepsilon }^{4}$.
Therefore, Therefore, the SNR corresponding to the theoretical variance of the data sequence obtained by Prop.2 is:
\begin{equation}\label{Eqn18}
\begin{aligned}
    \text{SN}{{\text{R}}_{2}}&=10{{\log }_{10}}\left[ \frac{\operatorname{var}\left( s_{X,k}^{2} \right)}{\operatorname{var}\left( s_{\varepsilon,k}^{2}  \right)} \right]
    \\&\approx 2\text{SN}{{\text{R}}_\textup{raw}}+10{{\log }_{10}}\left[ \frac{1}{{{T}_{\text{s}}}}\int_{-{n{{T}_{\text{s}}}}/2}^{{n{{T}_{\text{s}}}}/2}{{\rho }{{\left( \tau  \right)}^{2}}}\text{d}\tau  \right].
\end{aligned}
\end{equation}
The integral term measures the correlation strength of  $X\left( t \right)$ within interval $\tau \in\left[ -n{{T}_{\text{s}}}/2,n{{T}_{\text{s}}}/2 \right]$, capturing the variation of the autocovariance function over the interval.

As shown in \eqref{Eqn18}, since effective ambient fluctuations are modeled as colored noise, while measurement noise is typically modeled as white noise, $\text{SNR}_2$ is significantly improved compared to $\text{SNR}_\textup{raw}$, highlighting the effectiveness of sliding-window variance statistics in enhancing signal clarity. The value of $\text{SNR}_2$ depends on the autocorrelation function $\rho(\tau)$ of the ambient fluctuations. When $\rho(\tau)$ decays slowly, adjacent samples exhibit stronger temporal correlation, and the SNR improvement provided by sliding-window statistics becomes more pronounced. Under rapidly decaying autocorrelation conditions, improvements in SNR can still be achieved by increasing the window size $W$ or reducing the sampling period ${T}_{\text{s}}$.

For example, electrical quantity fluctuations are modeled using an O-U process with added GWN as measurement noise. The original data has an SNR of 0 dB, indicating low data quality. In this study, the decay rate $\alpha$ of the O-U process is set to 1, representing the fastest autocorrelation decay among the cases examined in \cite{Adeen2021}. This selection reflects the least favorable autocorrelation scenario, providing a stringent test of the method’s robustness. Fig.~\ref{Fig4_SNR} shows that stochastic characteristic-based methods significantly improve SNR, especially the Prop.2 method. Sensitivity identification using variance and covariance within a sliding window effectively enhances the system's SNR.

\begin{figure}[!htbp]
	\centering	\includegraphics[width=0.7\columnwidth]{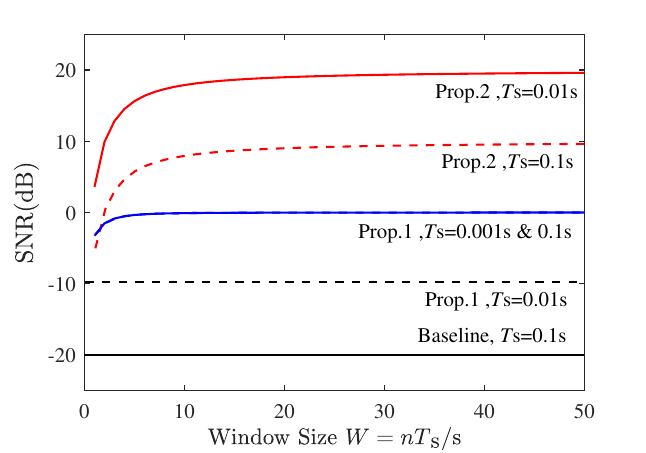}
	\caption{SNR of different identification methods. The baseline method refers to the temporal increments based approach as in \eqref{Eqn4}. Prop.1 denotes the mean-based identification method using a sliding window as in \eqref{Eqn9}, and Prop.2 refers to the variance-based identification method using sample variance and covariance as in \eqref{Eqn10}. The excitation is modeled by an O-U process with decay rate $\alpha=1$, diffusion coefficient $b=\sqrt{2}$, variance $\sigma^2=1$, and white noise with variance 1 is added as measurement noise.}
	\label{Fig4_SNR}
\end{figure}

\begin{figure}[!htbp]
	\centering	\includegraphics[width=1.0\columnwidth]{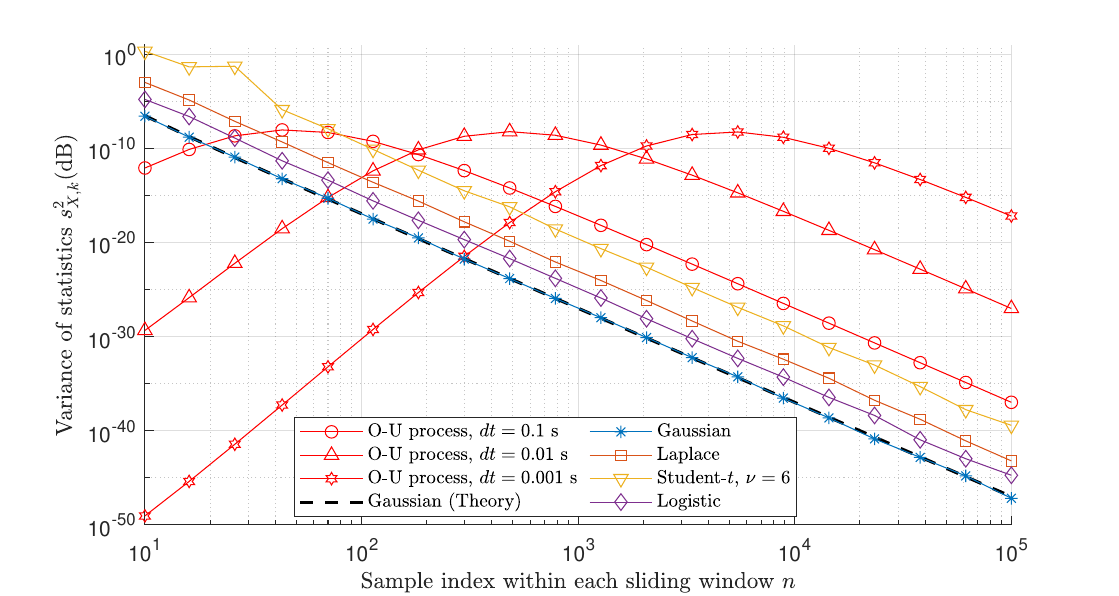}
	\caption{Variance of sample statistics of Prop.2 method for O-U process (with varying sampling intervals) and noice with different distributions. The variance gap between the O-U process and the noise sequences effectively reflects the SNR of the system within the sliding window.}
	\label{Figadd_SNR}
\end{figure}

{The theoretical analysis of SNR improvement given above is derived under the assumption of GWN. Fig.~\ref{Figadd_SNR} compares the performance of Prop.2 method under O-U process excitation with Gaussian and non-Gaussian measurement noise.
As shown in Fig.~\ref{Figadd_SNR}, although $s^2_{\epsilon,k}$ of non-Gaussian noise is indeed higher than Gaussian noise, this increase remains bounded. The $s^2_{X,k}$ computed from O-U processes initially increases and then gradually decreases as the sliding window size grows, exhibiting a much slower decay compared to uncorrelated Gaussian or non-Gaussian noise. As the window size increases, $s^2_{X,k}$ show larger variances than those from uncorrelated noise.  This comparison validates that the proposed identification method can effectively improve the SNR and separate signal from noise, demonstrating robust and reliable performance even under non-Gaussian noise conditions.

The robustness and applicability of the proposed method under non-Gaussian noise conditions are further discussed and validated in Section V.}

\subsubsection{Robustness under Asynchronous Measurements}
When asynchronous measurement issues occur, assume a delay of $m$ sampling periods, i.e., a time shift $\tau=mT_{\text{s}}$, $m>0$. The theoretical value of the measurement data is ${{Y}_{k}}$, while the actual measurement is ${{\tilde{Y}}_{k}}$. The variance-based deviation ratio can be quantified as
\begin{equation}\label{Eqn19}
    {{\epsilon }_{r}}=\frac{\operatorname{var}\left( {{{\tilde{Y}}}_{k}}-{{Y}_{k}} \right)}{\operatorname{var}\left( {{Y}_{k}} \right)}=2-\frac{2\operatorname{cov}\left( {{Y}_{k}},{{{\tilde{Y}}}_{k}} \right)}{\operatorname{var}\left( {{Y}_{k}} \right)}.
\end{equation}
This deviation ratio reflects the degree of distortion in the input features caused by measurement misalignment. The following compares the deviation ratio of three identification methods under asynchronous measurement conditions.

For the baseline method, the deviation ratio between ${{\tilde{Y}}_{k}}={{X}_{k+m+1}}-{{X}_{k+m}}$ and ${{Y}_{k}}={{X}_{k+1}}-{{X}_{k}}$ is 
\begin{equation}\label{Eqn20}
    {{\epsilon }_{r}}=2-\frac{2\rho \left( m{{T}_{\text{s}}} \right)-\rho \left( \left( m-1 \right){{T}_{\text{s}}} \right)-\rho \left( \left( m+1 \right){{T}_{\text{s}}} \right)}{1-\rho \left( {{T}_{\text{s}}} \right)}.
\end{equation}

For identification based on statistical characteristics, the true values of the random variables within the sliding window are $\bm{X}={{\left[ {{X}_{k}},{{X}_{k+1}},\cdots ,{{X}_{k+n-1}} \right]}^\text{T}}$, and the measured values are $\tilde{\bm{X}}={{\left[ {{X}_{k+m}},{{X}_{k+m+1}},\cdots ,{{X}_{k+m+n-1}} \right]}^\text{T}}$. The covariance matrix between $\bm{X}$ and $\tilde{\bm{X}}$ is denoted as $\mathbf{{{\tilde{C}}_{X}}}$. Since the elements of the matrix satisfy ${{\tilde{C}}_{X\left( i,j \right)}}=\operatorname{cov}\left( {{X}_{k+i-1}},{{X}_{k+m+j-1}} \right),i,j=1,2,\cdots ,n$, $\mathbf{{{\tilde{C}}_{X}}}$ is no longer a symmetric Toeplitz matrix. The relationship between the elements in $\mathbf{{{\tilde{C}}_{X}}}$ and $\rho \left( \tau  \right) $ is 
\begin{equation*}
    \tilde{C}_{X(i,j)} = \sigma_X^2 \cdot \rho\left( |i - j+m| T_s \right), \quad i, j = 1, 2, \dots, n
\end{equation*}


For the mean-based method (Prop.1), the theoretical value for each sliding window is ${{Y}_{k}}={{X}_{k}}-{{\bar{X}}_{\text{w,}k}}$, and the actual measured value should be ${{\tilde{Y}}_{k}}={{Y}_{k+m}}$. The covariance between ${{Y}_{k}}$ and ${{\tilde{Y}}_{k}}$ is given by 
\begin{equation}\label{Eqn21}
    \begin{aligned}
        \operatorname{cov}\left( {{Y}_{k}},{{{\tilde{Y}}}_{k}} \right)&={{\tilde{C}}_{X(1,1)}}-\frac{1}{n}\sum\nolimits_i {{{{\tilde{C}}}_{X(i,1)}}} 
        \\&
        -\frac{1}{n}\sum\nolimits_j {{{{\tilde{C}}}_{X(1,j)}}} 
        +\frac{1}{{{n}^{2}}}\sum\nolimits_{i,j} {{{{\tilde{C}}}_{X(i,j)}}}.
    \end{aligned}
\end{equation}

For the variance-based method (Prop.2), the theoretical value for each sliding window is ${{Y}_{k}}=s_{X,k}^{2}$, and the actual measured value should be ${{\tilde{Y}}_{k}}={{Y}_{k+m}}=s_{X,k+m}^{2}$. The covariance between ${{Y}_{k}}$ and ${{\tilde{Y}}_{k}}$ is given by 
\begin{equation}\label{Eqn22}
        \operatorname{cov}\left( Y_k, \tilde{Y}_k \right) = \frac{2}{(n-1)^2} \left( 
        S_{1b} + \frac{1}{n^2} S_{2b} - \frac{1}{n} S_{3b} 
        \right) 
\end{equation}
where $S_{1b} = \sum_{i,j} \tilde{C}_{X(i,j)}^2$, $S_{2b} = \left( \sum_{i,j} \tilde{C}_{X(i,j)} \right)^2$, $S_{3b} = \sum_{i} \left( \sum_{j} \tilde{C}_{X(i,j)} \right)^2+\sum_{j} \left( \sum_{i} \tilde{C}_{X(i,j)} \right)^2$.

Taking the O-U process as an example, when asynchronous measurement issues occur in the signal, the deviation ratio for different identification methods is shown in Fig.~\ref{Fig5_Asynchronous}.
\begin{figure}[!htbp]
	\centering
	\subfloat[]{
    \includegraphics[trim=0.1cm 0cm 0.8cm 0cm, clip,width=0.48\columnwidth]{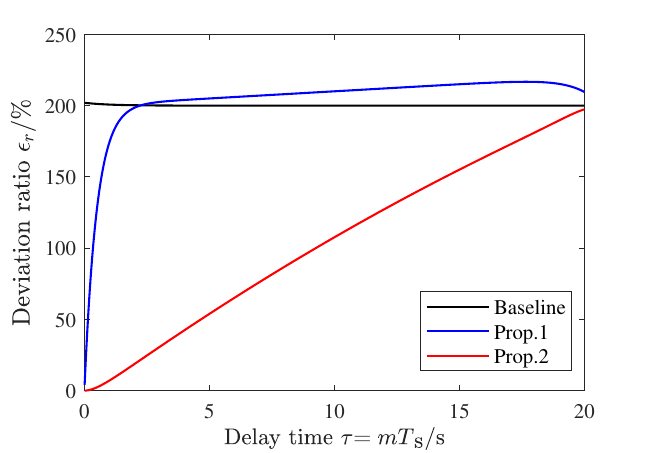}
		\label{Fig5a_Asynchronous}
	}
	\subfloat[]{
    \includegraphics[trim=0.1cm 0.1cm 0.8cm 2.7cm, clip,width=0.48\columnwidth]{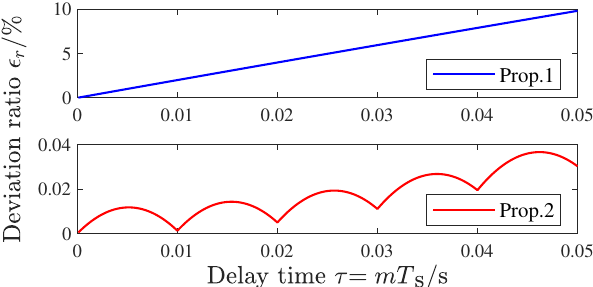}
		\label{Fig5b_Asynchronous}
	}    
	\caption{
    Variance-based deviation ratio of different identification methods under asynchronous measurement conditions. (a) shows the case where the delay $\tau$ is an integer multiple of $T_\text{s}$, and (b) shows the non-integer case. The baseline method is based on temporal increments as in \eqref{Eqn4}, while Prop.1 and Prop.2 use mean-based as in \eqref{Eqn9} and variance-based as in \eqref{Eqn10} identification under sliding windows, respectively. O-U process parameters and sampling settings are the same as in Fig.~\ref{Fig4_SNR}. The sampling period is ${{T}_{\text{s}}}=0.01\,\text{s}$, and the sliding window length is $W=20\,\text{s}$. The deviation ratio is defined by \eqref{Eqn19}.}
	\label{Fig5_Asynchronous}
\end{figure}

Fig.~\ref{Fig5_Asynchronous} demonstrates the excellent robustness of the method based on stochastic characteristics under asynchronous measurements, especially Prop.2 method. Even in the presence of time delays in asynchronous measurement signals, sensitivity identification using the variance and covariance under the sliding window can still provide accurate estimates, demonstrating the method's robustness against measurement inconsistencies.

\subsubsection{Handling of Data Collinearity}
The active and reactive power data may exhibit bivariate collinearity (the power correlation is modeled according to \cite{Ref28_adeen2021modeling}), with the correlation coefficient between the power data denoted as $\operatorname{corr}\left( P_k,Q_k \right)=r_{PQ}$, where $0\le r_{PQ}\le 1$. We assume that the variances $\sigma_P^2$ and $\sigma_Q^2$ are equal. The condition numbers of the power data for each identification method can be  computed to assess robustness under such collinearity.

Consider the $n\times m$ random matrix $\mathbf{A}$, formed by the measurement data column vectors of $m$ correlated random variables, which follow a Gaussian distribution. By centering $\mathbf{A}$ with column means, we obtain $\mathbf{A}_{\text{dec}}$. The 2-norm condition number $\kappa(\mathbf{A}_{\text{dec}})$ is the ratio of the largest to smallest singular value of $\mathbf{A}_{\text{dec}}$, depending on the correlation between its columns.

The squared singular values $\sigma_i^2$ of matrix $\mathbf{A}_{\text{dec}}$ are equal to the eigenvalues $\lambda_i$ of matrix ${\mathbf{A}_{\text{dec}}^{\text{T}}{\mathbf{A}_{\text{dec}}}}$. Thus, the square of $\kappa \left( \mathbf{A}_{\text{dec}} \right)$ equals the condition number of the covariance matrix $\operatorname{cov}(\mathbf{A})=\frac{1}{n-1}\mathbf{A}_{\text{dec}}^{\text{T}}{\mathbf{A}_{\text{dec}}}$ \cite{Ref29_akemann2011oxford}. When the $m$ column random fluctuations have equal variance, the square of $\kappa \left( \mathbf{A}_{\text{dec}} \right)$ equals the condition number of the correlation matrix $\operatorname{corr}\left( \mathbf{A}_{\text{dec}} \right)$.

For the baseline method, the covariance between $P_{k+1}-P_k$ and $Q_{k+1}-Q_k$ is
\begin{equation}\label{Eqn23}
    \operatorname{cov}\left( P_{k+1} - P_k, Q_{k+1} - Q_k \right) = 2 r_{PQ} \sigma_P \sigma_Q \left( 1 - \rho(T_s) \right)
\end{equation}
Combining \eqref{Eqn12} and \eqref{Eqn23}, the correlation coefficient matrix of $\mathbf{A}_k=\left[ {{P}_{k+1}}-{{P}_{k}},{{Q}_{k+1}}-{{Q}_{k}} \right]$ is given by
\begin{equation}\label{Eqn24}
    \text{corr}\left( \mathbf{A} \right)=\left[ 
    \begin{matrix}
       1 & {{r}_{PQ}}  \\
       {{r}_{PQ}} & 1  \\
    \end{matrix} 
    \right].
\end{equation}
The theoretical condition number of $\mathbf{A}_\text{dec}$ is
\begin{equation}\label{Eqn25}
    {{\kappa }_{0}}\left( {\mathbf{A}_{\text{dec}}} \right)=\sqrt{\frac{1+{{r}_{PQ}}}{1-{{r}_{PQ}}}}.
\end{equation}

For identification based on statistical characteristics, the covariance matrix between $\bm{P}={{\left[ {{P}_{k}},{{P}_{k+1}},\cdots ,{{P}_{k+n-1}} \right]}^{\text{T}}}$ and $\bm{Q}={{\left[ {{Q}_{k}},{{Q}_{k+1}},\cdots ,{\bm{Q}_{k+n-1}} \right]}^{\text{T}}}$ is denoted as $\mathbf{C_{PQ}}$. 
According to the modeling method in \cite{Ref28_adeen2021modeling}, when the normalized autocovariance functions ${{\rho }_{P}}\left( \tau  \right)$ and ${{\rho }_{Q}}\left( \tau \right)$ are identical, the normalized cross-covariance function between $\bm{P}$ and $\bm{Q}$ is the same as ${{\rho }_{P}}\left( \tau  \right)$ and ${{\rho }_{Q}}\left( \tau \right)$. Therefore, the relationship between the elements in $\mathbf{C_{PQ}}$  and $\rho \left( \tau  \right) $ is 
\begin{equation*}
    \tilde{C}_{X(i,j)} = r_{PQ}\sigma_P\sigma_Q \cdot \rho\left( |i - j| T_s \right), \quad i, j = 1, 2, \dots, n
\end{equation*}


For the mean-based method (Prop.1), the covariance between ${{P}_{k}}-{{\bar{P}}_{\text{w,}k}}$ and
${{Q}_{k}}-{{\bar{Q}}_{\text{w,}k}}$ is given by
\begin{equation}\label{Eqn26}
    \begin{aligned}
        \operatorname{cov} \big( P_k - \bar{P}_{w,k} &, Q_k - \bar{Q}_{w,k} \big)  = C_{PQ(1,1)} 
         \\&- \frac{2}{n} \sum\nolimits_i C_{PQ(i,1)} 
         + \frac{1}{n^2} \sum\nolimits_{i,j} C_{PQ(i,j)}.
    \end{aligned}    
\end{equation}
Combining \eqref{Eqn14} and \eqref{Eqn26}, the correlation coefficient matrix of $\mathbf{A}_k=\left[ {{P}_{k}}-{{{\bar{P}}}_{\text{w,}k}},{{Q}_{k}}-{{{\bar{Q}}}_{\text{w,}k}} \right]$ is the same as in \eqref{Eqn24}. The condition number ${{\kappa }_{1}}\left( {\mathbf{A}_{\text{dec}}} \right)={{\kappa }_{0}}\left( {\mathbf{A}_{\text{dec}}} \right)$.

For the variance-based method (Prop.2), following a similar approach, we derive the theoretical values of $\operatorname{cov}\left( s_{P,k}^{2},s_{Q,k}^{2} \right)$, $\operatorname{cov}\left( s_{P,k}^{2},s_{PQ,k} \right)$, $\operatorname{cov}\left( s_{Q,k}^{2},s_{PQ,k} \right)$ and $\operatorname{var}\left( s_{PQ,k} \right)$. By combining these theoretical values, the correlation coefficient matrix of $\mathbf{A}_k=\left[ s_{P,k}^{2},s_{Q,k}^{2},{{s}_{PQ,k}} \right]$ is
\begin{equation}\label{Eqn28}
    \text{corr}\left( \mathbf{A} \right)=\left[ 
    \begin{matrix}
       1 & r_{PQ}^{2} & \sqrt{\frac{2}{1+r_{PQ}^{2}}}{{r}_{PQ}}  \\
       r_{PQ}^{2} & 1 & \sqrt{\frac{2}{1+r_{PQ}^{2}}}{{r}_{PQ}}  \\
       \sqrt{\frac{2}{1+r_{PQ}^{2}}}{{r}_{PQ}} & \sqrt{\frac{2}{1+r_{PQ}^{2}}}{{r}_{PQ}} & 1 \\
    \end{matrix} 
    \right].
\end{equation}
The theoretical condition number of $\mathbf{A}_\text{dec}$ is
\begin{equation}\label{Eqn28}
    \resizebox{1.0\columnwidth}{!}{$
        {{\kappa }_{2}}\left( {\mathbf{A}_{\text{dec}}} \right)=\sqrt{\frac{\sqrt{r_{PQ}^{2}+1}\left( r_{PQ}^{2}+2 \right)+{{r}_{PQ}}\sqrt{r_{PQ}^{4}+r_{PQ}^{2}+16}}{\sqrt{r_{PQ}^{2}+1}\left( r_{PQ}^{2}+2 \right)-{{r}_{PQ}}\sqrt{r_{PQ}^{4}+r_{PQ}^{2}+16}}}.
    $}
\end{equation}

The theoretical condition number of the data matrix based on different identification methods is shown in Fig.~\ref{Fig6_Collinearity}, with variations in data correlation. As shown in Fig.~\ref{Fig6_Collinearity}, when the correlation of $P$ and $Q$ is low, the proposed identification methods based on mean and variance do not significantly exacerbate the data collinearity. However, when correlation is extremely high, the variance-based method leads to a noticeable increase in data collinearity.
\begin{figure}[!htbp]
	\centering
    \includegraphics[width=0.7\columnwidth]{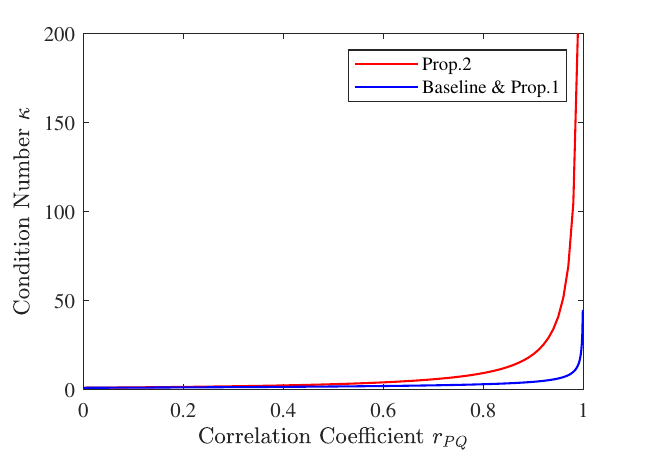}
	\caption{
    Condition numbers of different identification methods under various correlation levels between $P$ and $Q$. The baseline method uses temporal increments as in \eqref{Eqn4}, while Prop.1 and Prop.2 use statistical characteristics based on mean as in \eqref{Eqn9} and variance as in \eqref{Eqn10} respectively. The higher the condition number, the more severe the collinearity problem.}
	\label{Fig6_Collinearity}
\end{figure}

It is worth noting that although the variance-based method (Prop.2) results in a slightly higher condition number under extreme correlation, it does not imply degraded robustness. As derived in Section III-C, the final parameter error is jointly affected by both the condition number and the relative perturbations in the input and output data. Since the proposed methods significantly reduce the effect of random noise (as discussed in Section III-B.1), the overall estimation error can still be suppressed effectively even when the condition number is not minimized. This highlights that robustness should not be evaluated solely by structural metrics like condition number, but also by considering the interaction between data structure and noise sensitivity. Therefore, the robustness of the proposed methods is better interpreted from the full error bound perspective, rather than solely from the viewpoint of matrix conditioning.

\subsection{Method Selection Based on Error Bounds}

The sensitivity identification error is affected by SNR, asynchronous measurements, and data collinearity. Sensitivity identification can be regarded as an $\mathbf{A}\bm{\Theta}=\mathbf{B}$ type identification. The data and identification objectives for different identification methods are shown in Table I.

Due to varying data expectations across different sensitivity identification methods (the perturbation sequence $s_{\varepsilon,k}^{2}$ in Prop.2 follows a chi-square distribution), de-meaning is necessary before calculating the error bound. According to the law of large numbers, with a large sample size, the sample mean approximates the expected value. This operation removes systematic mean shifts, ensuring that the identification accuracy reflects the method’s robustness to fluctuations, rather than being influenced by the signal magnitude or units.

The error upper bound can be effectively controlled through the condition number criteria \cite{Baboulin2009} applied to the decentralized data, as 
\begin{equation}\label{Eqn29}
\resizebox{1.0\columnwidth}{!}{$
    \frac{\|\hat{\bm{\Theta}} - \bm{\Theta}\|_{\text{F}}}{\|\bm{\Theta}\|_{\text{F}}} \le \kappa_{\text{F}}(\mathbf{A}_{\text{dec}}) \left( \frac{\| \tilde{\mathbf{A}}_{\text{dec}} - \mathbf{A}_{\text{dec}} \|_{\text{F}}}{\|\mathbf{A}_{\text{dec}}\|_{\text{F}}} + \frac{\| \tilde{\mathbf{B}}_{\text{dec}} - \mathbf{B}_{\text{dec}} \|_{\text{F}}}{\|\mathbf{B}_{\text{dec}}\|_{\text{F}}} \right)
    $}
\end{equation}
where, ${{\left\| \cdot  \right\|}_{\text{F}}}$ represents the Frobenius norm (F-norm). The F-norm of a decentralized random sequence is proportional to the variance of the sequence. $\kappa_{\text{F}} \left(  \cdot \right)$ represents the condition number of the matrix under the F-norm.
$\hat{\boldsymbol{\Theta}}$ denotes the estimated value of MSP.
${{\tilde{\mathbf A}}_{\text{dec}}}$ and ${{\tilde{\mathbf{B}}}_{\text{dec}}}$ represents the actual measured value of ${\mathbf A}$ and ${\mathbf{B}}$ after decentralization, respectively. Since $\boldsymbol{\Theta}_2 = \mathcal{T}(\boldsymbol{\Theta}_1)$ (as in Table I) is a nonlinear transformation, the estimation error of Prop.2 should be evaluated via the mapped parameters. Based on first-order approximation, $\|\hat{\boldsymbol{\Theta}}_2 - \boldsymbol{\Theta}_2\|_\textup{F} \leq \|J_{\mathcal{T}}(\boldsymbol{\Theta}_1)\| \cdot \|\hat{\boldsymbol{\Theta}}_1 - \boldsymbol{\Theta}_1\|_\textup{F}$, where $J_{\mathcal{T}}(\boldsymbol{\Theta}_1)$ is the Jacobian of $\mathcal{T}(\cdot)$ evaluated at $\boldsymbol{\Theta}_1$.

The SNR improvement and the deviation ratio under asynchronous measurements are related to the F-norms of the matrices $\mathbf A$ and $\mathbf{B}$. Since $ \mathbf{A} \in \mathbb{R}^{k \times 2}$, $\kappa_{\text{F}} \left( \mathbf{A}_{\text{dec}} \right)$ is positively correlated with $\kappa \left( \mathbf{A}_{\text{dec}} \right)$ and the column correlation of $\mathbf{A}$, with $\kappa\left( \mathbf{A}_{\text{dec}} \right) \leq \kappa_{\text{F}}\left( \mathbf{A}_{\text{dec}} \right) \leq 2 \cdot \kappa\left( \mathbf{A}_{\text{dec}} \right)$ \cite{Golub2013}.
Based on the condition number criteria, the error upper bound of the solution in each method can be estimated. Qualitatively, under low SNR or asynchronous conditions, the variance-based method is preferred, while under high collinearity, the mean-based method is more suitable.

\subsection{Selection of Sliding Window Parameters}
For the proposed method, the window size $W$ and the sliding step $S$ are key factors affecting both computational accuracy and efficiency. 

To maximize data utilization and robustness, $S$ is typically set to $T_\text{s}$, moving the window one step at a time.

For $W$, the goal is to optimize SNR improvement. As shown in Fig.~\ref{Fig4_SNR}, while longer windows can enhance estimation accuracy, reducing the data load per calculation is key to improving efficiency. First, the signal’s autocovariance function is estimated and converted into a covariance matrix. Using  \eqref{Eqn16}, SNR improvement is evaluated for various window lengths. In practice, $W$ is selected based on the signal's autocovariance decay time constant $\tau _c$, with $W=5\tau _c\sim10\tau _c$. If the identification results don't converge, $W$ can be further increased or the system’s sampling period $T_\text{s}$ reduced.

\section{Simulation Verification}
\subsection{Simulation Validation of Method Effectiveness}\label{sec:MATLAB}
\subsubsection{Case Setup}
The case model used for the simulation is shown in Fig.~\ref{Fig1_Thevenin_equivalent}. The simulation is conducted in the MATLAB environment. The test case parameters are as follows: the source voltage is $E=270\,\text{kV}$, the impedance is $Z=R+\text{j}X=\left( 20+\text{j}50 \right)\,\Omega$. The active and reactive power of the stochastic load are defined by a bivariate correlated O-U process \cite{Ref28_adeen2021modeling} as
\begin{equation}
\begin{cases}
\begin{aligned}
  & P\left( t \right)=\left( {{P}_{0}}+{{\eta }_{P}}\left( t \right) \right){{\left( \left| V \right|\left( t \right)/\left| {{V}_{0}} \right| \right)}^{{{\gamma }_{P}}}} \\ 
 & Q\left( t \right)=\left( {{Q}_{0}}+{{\eta }_{Q}}\left( t \right) \right){{\left( \left| V \right|\left( t \right)/\left| {{V}_{0}} \right| \right)}^{{{\gamma }_{Q}}}} \\ 
 & {{{\dot{\eta }}}_{P}}\left( t \right)=-{{\alpha }_{P}}{{\eta }_{P}}\left( t \right)+{{b}_{P}}{{\xi }_{P}}\left( t \right) \\ 
 & {{{\dot{\eta }}}_{Q}}\left( t \right)=-{{\alpha }_{Q}}{{\eta }_{Q}}\left( t \right)+{{b}_{Q}}\left( {{r}_{PQ}}{{\xi }_{P}}\left( t \right)+\sqrt{1-r_{PQ}^{2}}{{\xi }_{Q}}\left( t \right) \right)
\end{aligned}
\end{cases}
\end{equation}
where the mathematical expectations of power are $P_0=50\,\text{MW}$ and  $Q_0=50\,\text{MVar}$, $\left| V \right|\left( t \right)$ is the magnitude of the bus voltage at the load side, $\left| {{V}_{0}} \right|$ is the initial value of the voltage magnitude at time $t=0$. $\gamma_P$ and $\gamma_Q$ represent the voltage-power correlation at the load side. $\gamma _P=\gamma _Q=0$ indicates a constant power load (CPL), and $\gamma _P=\gamma _Q=2$ indicates a constant impedance load (CIL).

$\eta_P\left( t \right)$ and $\eta_Q\left( t \right)$  are modeled as bivariate correlated zero-mean O-U processes. The decay rate is $\alpha _P=\alpha _Q=1$. The diffusion coefficient is $b_P=b_Q=\sqrt{2}$. The variance is $\sigma _{{{\eta }_{P}}}^{2}=\sigma _{{{\eta }_{Q}}}^{2}=1$. ${{\xi }_{P}}\left( t \right)$ and ${{\xi }_{Q}}\left( t \right)$ are independent standard GWN, with parameter $r_{PQ}$ indicating their correlation. When $r_{PQ}=0$, the fluctuations in active and reactive power are uncorrelated.

\subsubsection{Effectiveness for Various Load Types}
To verify the accuracy of the proposed TEP identification method under different load types, simulations are conducted with CPL and CIL. The method’s performance under varying load conditions is evaluated through simulation comparisons.

In addition to the proposed method, we compare ordinary LS, ridge regression \cite{Ref24_tang2024adaptive},\cite{Ref25_liang2023temporally}, and TLS \cite{Ref23_silva2020data} to assess the performance of different methods in TEP identification. The violin plot in Fig.~\ref{Fig_7} shows the probability distribution of TEP identification results for each method, providing an intuitive comparison of the accuracy and stability of the identification results under varying load conditions.

\begin{figure}[!htbp] 
    \centering
    \subfloat[$\hat{E}_\text{th}/E_\text{th}$ with CPL]{
		\includegraphics[trim=0.6cm 0.3cm 1.3cm 0.7cm, clip, width=0.48\columnwidth]{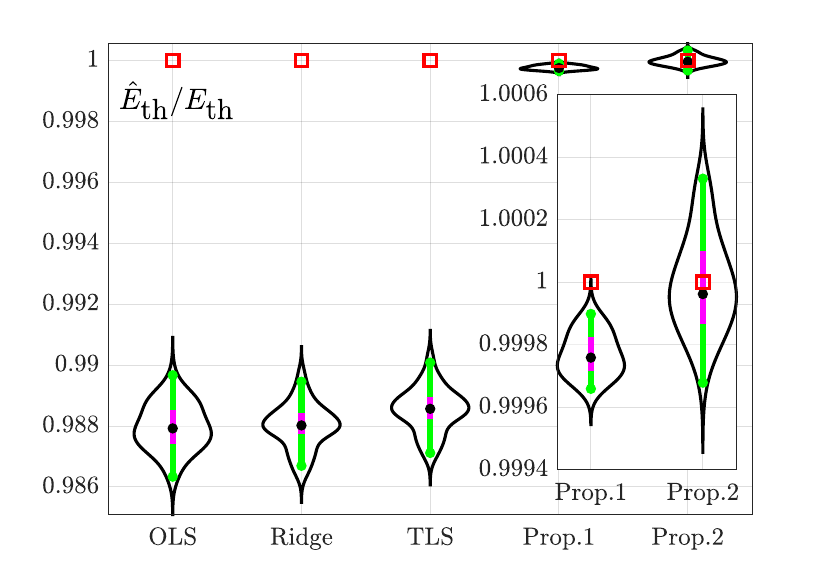}}
	\subfloat[$\hat{E}_\text{th}/E_\text{th}$ with CIL]{
		\includegraphics[trim=0.6cm 0.3cm 1.3cm 0.7cm, clip, width=0.48\columnwidth]{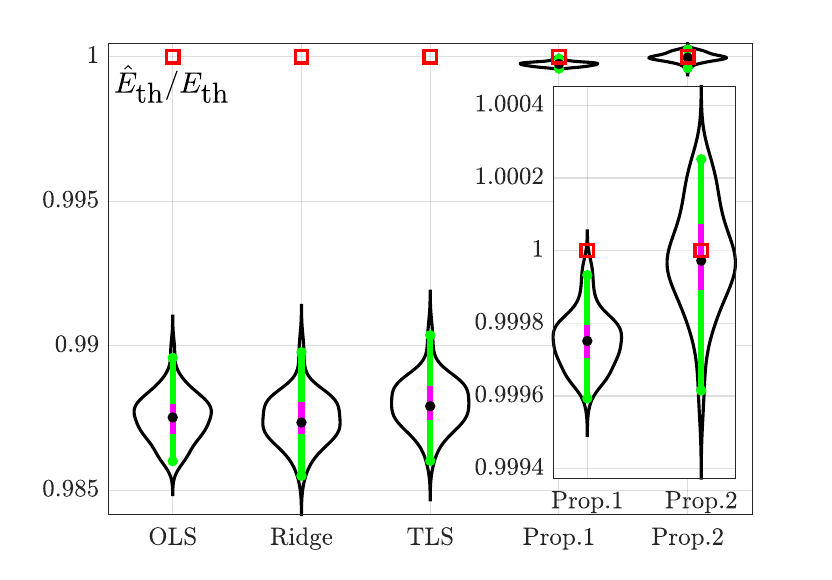}} \\
    \subfloat[$\hat{R}_\text{th}/R_\text{th}$ with CPL]{
		\includegraphics[trim=1.1cm 0.3cm 1.3cm 0.7cm, clip, width=0.48\columnwidth]{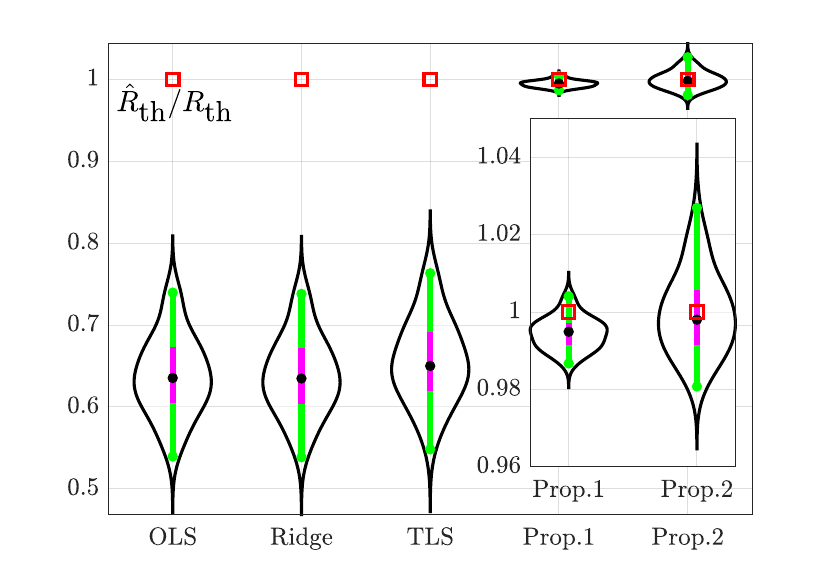}}
    \subfloat[$\hat{R}_\text{th}/R_\text{th}$ with CIL]{
		\includegraphics[trim=1.1cm 0.3cm 1.3cm 0.7cm, clip, width=0.48\columnwidth]{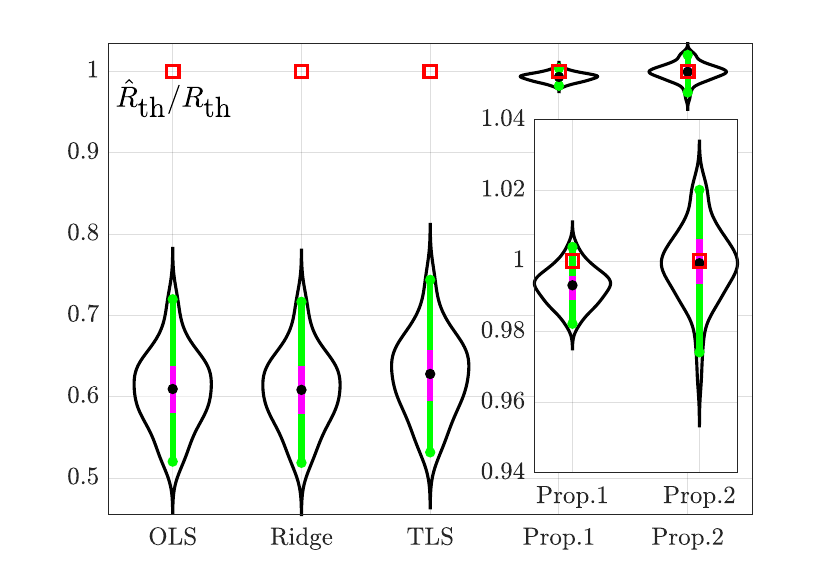}}\\ 
	\subfloat[$\hat{X}_\text{th}/X_\text{th}$ with CPL]{
		\includegraphics[trim=1.1cm 0.3cm 1.3cm 0.7cm, clip, width=0.48\columnwidth]{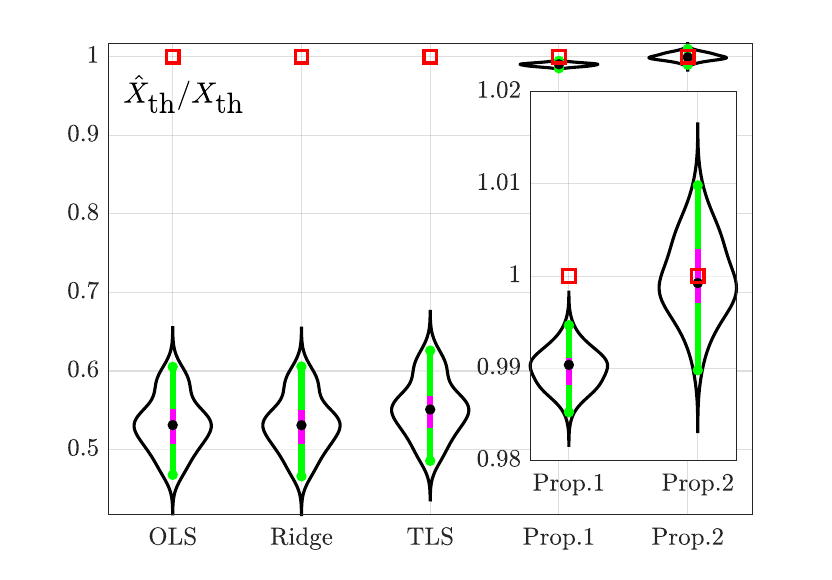}}
    \subfloat[$\hat{X}_\text{th}/X_\text{th}$ with CIL]{
		\includegraphics[trim=1.1cm 0.3cm 1.3cm 0.7cm, clip, width=0.48\columnwidth]{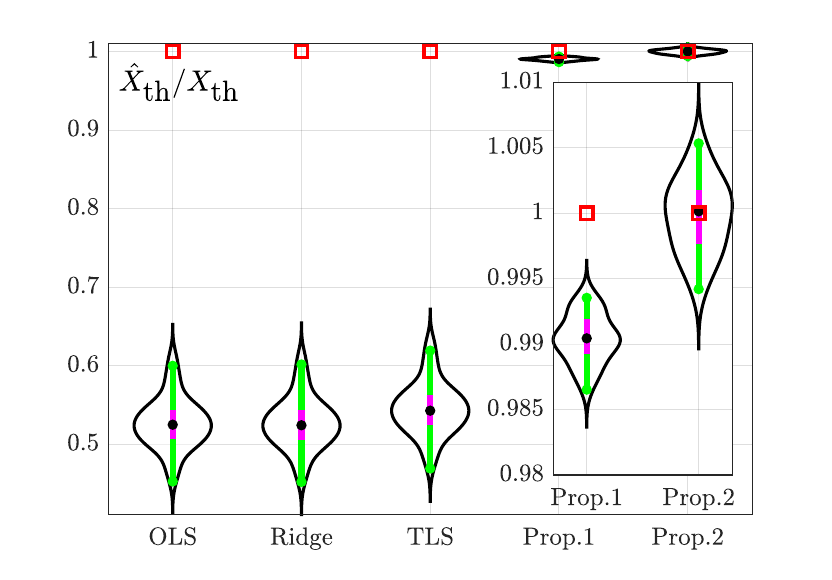}}
    \caption{Violin plots of TEP identification results under different load models. In the violin plots, the black solid line contour illustrates the probability distribution of results from each method (obtained via kernel density estimation), and the red box represents the theoretical true values of the parameters. The black solid circle marks the median of the identification results, while the hollow square represents the theoretical values. The magenta bar indicates the interquartile range (IQR), i.e., the difference between the 75th and 25th percentiles of the data. The green solid circles represent the nearest upper and lower bound values, which are the maximum observation less than or equal to the third quartile plus 1.5×IQR and the minimum observation greater than or equal to the first quartile minus 1.5×IQR, respectively\cite{Ref32_bizzarri2024inertia}.
    }\label{Fig_7}
\end{figure}

In Fig.~\ref{Fig_7}, the horizontal axis represents the different methods employed, including literature methods (OLS, Ridge, TLS) and the proposed methods (Prop.1 mean-based and Prop.2 variance-based). The correlation coefficients $r_{PQ}=0.2$. The noise variance in $\left| V \right|$, $\left| I \right|$, $P$, $Q$ are set to 1\% of the signal variance, corresponding to a SNR of 20 dB. The simulation is performed with 50 independent Monte Carlo experiments, each simulating for 2 minutes. The sampling period $T_{\text{s}}=0.01\,\text{s}$ and the sliding window size $W=5\,\text{s}$. The inset of Fig.~\ref{Fig_7} zooms in on the distribution of results from the proposed methods (Prop.1 mean-based and Prop.2 variance-based), allowing for a more detailed comparison of their accuracy and bias.

The simulation results demonstrate that the proposed methods based on stochastic characteristics maintain strong unbiasedness and stability across different load types. Compared to existing methods, the estimates from the proposed methods, with median values (black solid points in Fig.~\ref{Fig_7}) closely aligning with the theoretical values (empty square markers), confirm their unbiased nature. Additionally, the narrow interquartile range (IQR) (magenta vertical bars) and the deviation between adjacent values (green solid points) within 2\% further validate the consistent stability of the methods.

Since the identification accuracy of the three TEPs is similar, only the results for $X_\text{th}$ are presented here for brevity.

\subsubsection{Effectiveness under Low Data Quality}
 To comprehensively evaluate the effectiveness of the proposed method under low data quality, we conduct simulations under low SNR, asynchronous measurements, and high data collinearity. The results demonstrate the robustness and accuracy of the proposed method in these scenarios.

 \begin{figure}[!htbp] 
    \centering
    \subfloat[Low SNR: SNR= 10 dB]{
		\includegraphics[trim=1cm 0.3cm 1.3cm 0.7cm, clip, width=0.48\columnwidth]{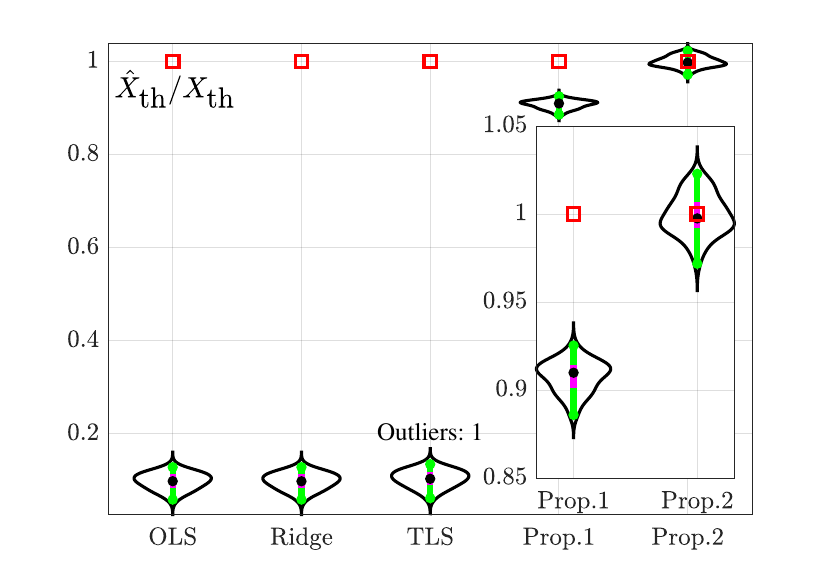}}
        \subfloat[Low SNR: SNR= 0 dB]{
		\includegraphics[trim=1cm 0.3cm 1.3cm 0.7cm, clip, width=0.48\columnwidth]{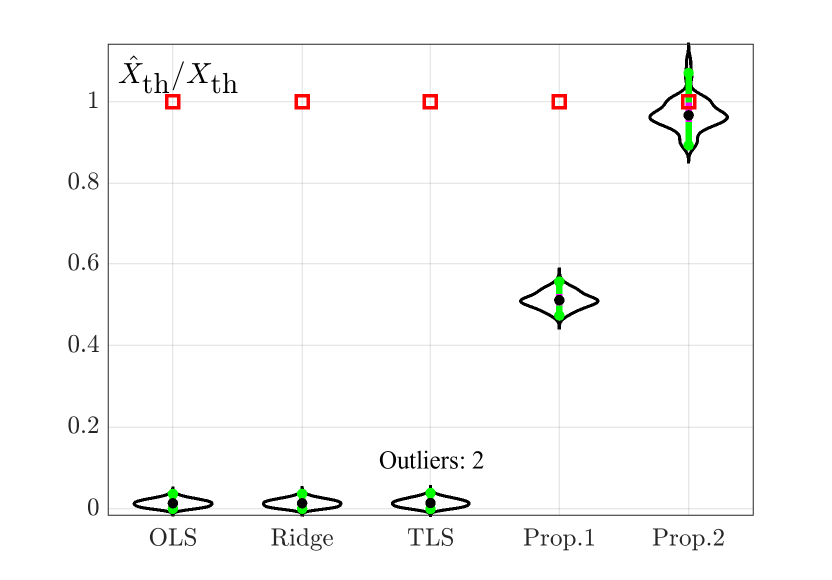}}\\
        \subfloat[Asynchronous:  $\tau=0.05~s$]{
        \includegraphics[trim=1cm 0.3cm 1.3cm 0.7cm, clip, width=0.48\columnwidth]{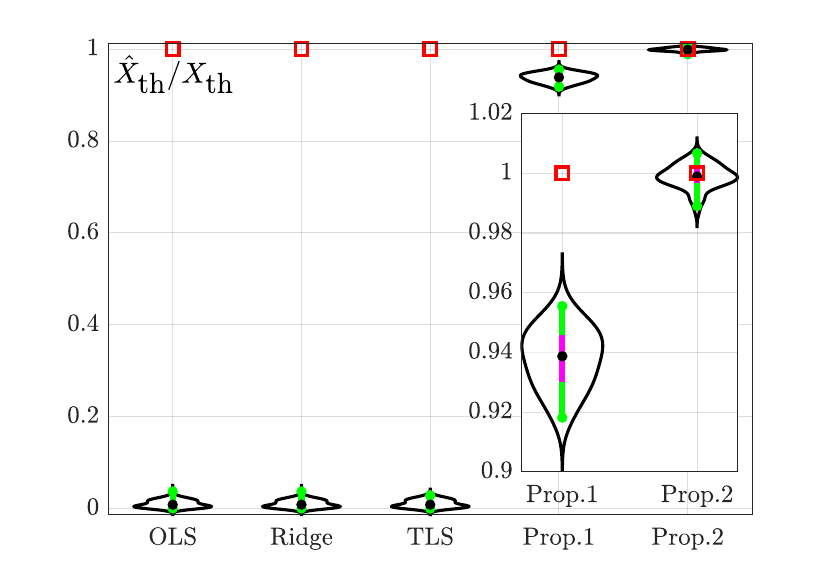}}
        \subfloat[Asynchronous: $\tau=0.1~s$]{
		\includegraphics[trim=1cm 0.3cm 1.3cm 0.7cm, clip, width=0.48\columnwidth]{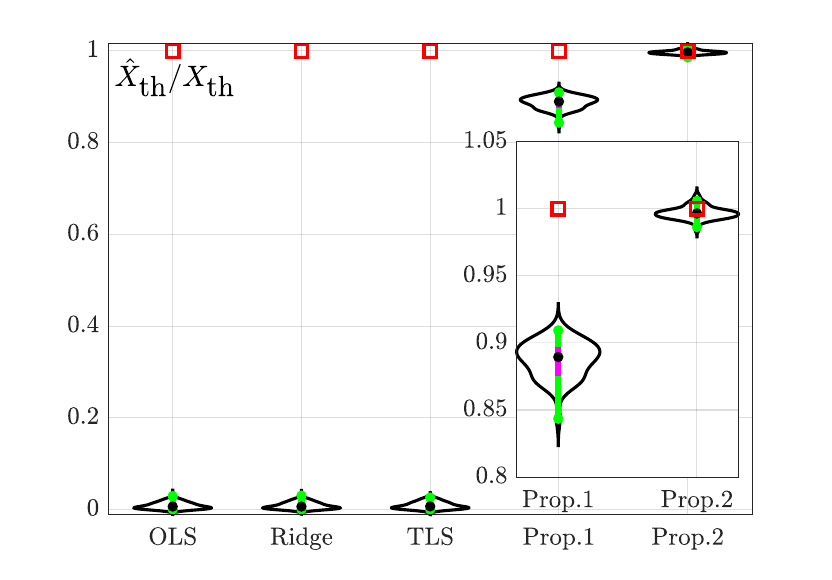}}\\
	\subfloat[High collinearity: $r_{PQ}=0.99$]{
		\includegraphics[trim=1cm 0.3cm 1.3cm 0.7cm, clip, width=0.48\columnwidth]{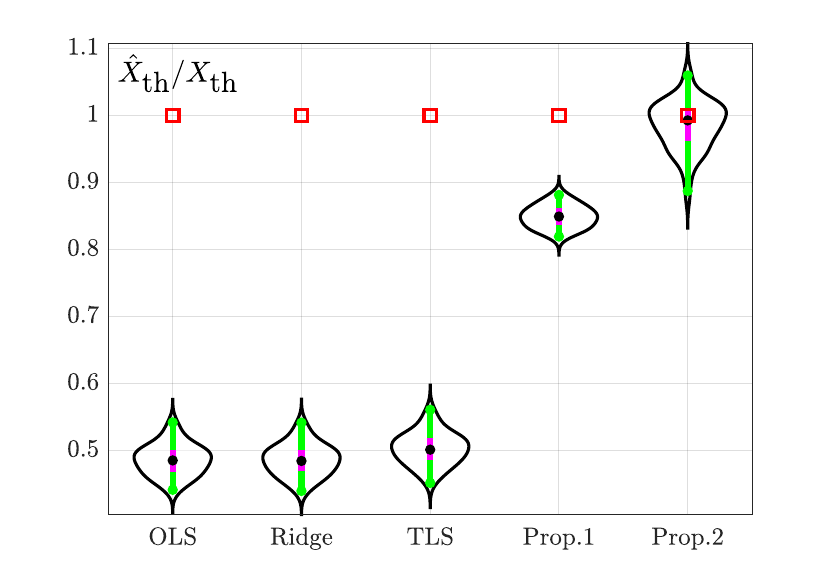}}
    \subfloat[High collinearity: $r_{PQ}=0.999$]{
		\includegraphics[trim=1cm 0.3cm 1.3cm 0.7cm, clip, width=0.48\columnwidth]{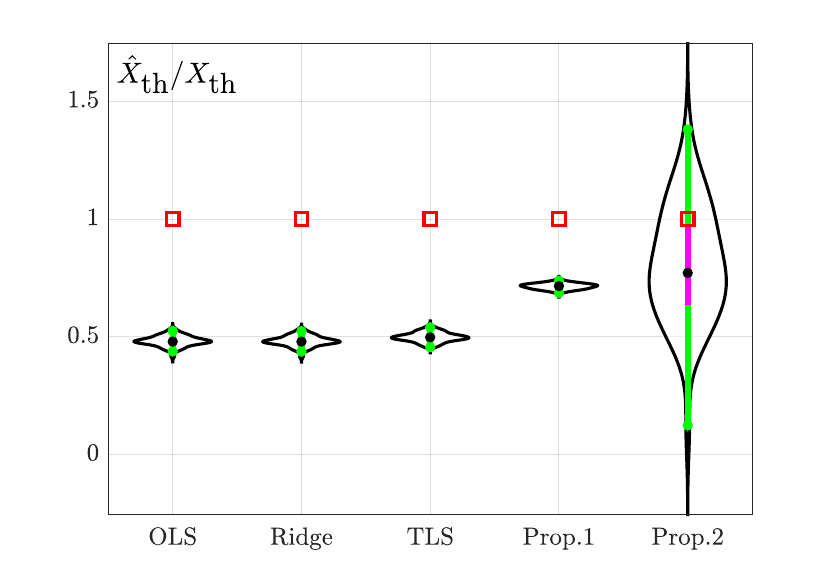}}\\ 
    \caption{Violin plots of $X_\text{th}$ identification results under different conditions. The load type is a CPL. $r_{PQ}$ represents the correlation coefficient. $\tau$ represents the delay time. All other simulation settings are consistent with Fig.~\ref{Fig_7}.}\label{Fig_8}
\end{figure}

\begin{figure}[!htbp] 
    \centering
    \subfloat[]{
		\includegraphics[trim=0.5cm 0.3cm 0.7cm 0.3cm, clip, width=0.48\columnwidth]{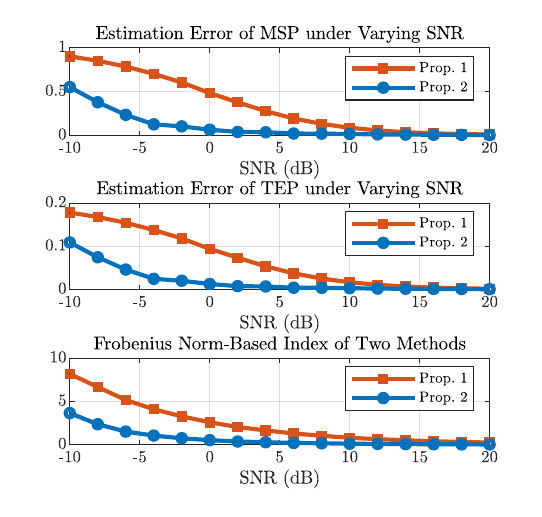}}\\
        \subfloat[]{
		\includegraphics[trim=0.6cm 0.3cm 0.7cm 0.3cm, clip, width=0.48\columnwidth]{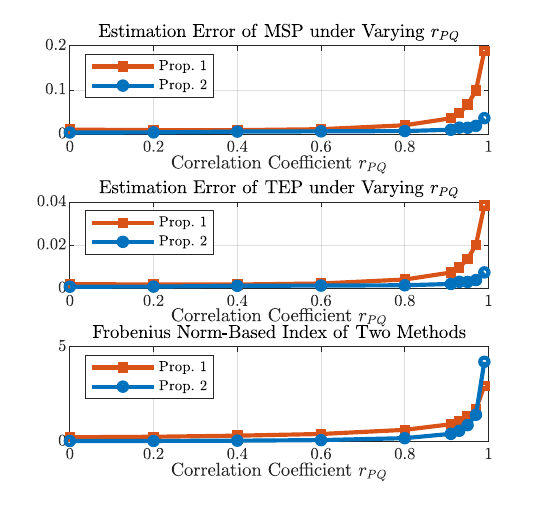}}
        \subfloat[]{
        \includegraphics[trim=0.6cm 0.3cm 0.7cm 0.3cm, clip, width=0.48\columnwidth]{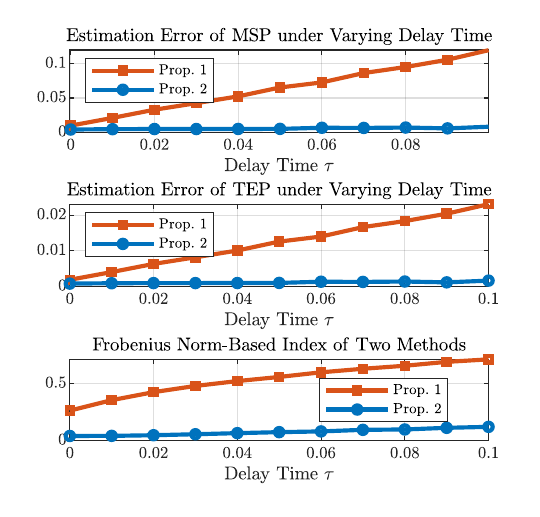}}
    \caption{Estimation relative error and Frobenius norm-based index of Prop.1 and Prop.2 methods under varying SNR, correlation coefficient and delay time conditions. Estimation errors are computed as relative Frobenius norms, defined by
    $\|\widehat{\boldsymbol{\Theta}} - \boldsymbol{\Theta}_{\mathrm{true}}\|_{\mathrm{F}} / \|\boldsymbol{\Theta}_{\mathrm{true}}\|_{\mathrm{F}}$,
    where $\widehat{\boldsymbol{\Theta}}$ and $\boldsymbol{\Theta}_{\mathrm{true}}$ denote the estimated and true TEP and MSP, respectively.}\label{Fig_err}
\end{figure}

For \textbf{low SNR conditions}, Fig.~\ref{Fig_8} (a)-(b) presents the violin plots of the different methods under low SNR conditions. The simulation results show that, even with reduced SNR, the proposed method consistently achieves high TEP identification accuracy, further demonstrating its robustness and advantages in noisy environments. As shown in  Fig.~\ref{Fig_err}(a), under varying SNR conditions, the MSP error, TEP error, and the theoretical Frobenius-norm-based index all decrease monotonically. Prop.2 consistently outperforms Prop.1 across all SNR levels, demonstrating stronger noise robustness.

For \textbf{asynchronous measurements conditions}, we introduced a 0.05~s delay and a 0.1~s delay in $V$ and $I$ measurements relative to $P$ and $Q$ to simulate time misalignment caused by asynchronous measurements. Fig.~\ref{Fig_8} (c)-(d) presents the violin plots of the methods under these conditions. The simulation results demonstrate that, despite the presence of measurement delays, the proposed method—particularly Prop.2—maintains high identification accuracy. The sliding window technique helps smooth out the effects of asynchronous measurements by averaging over multiple data points, making small delays negligible in the final estimation. As shown in  Fig.~\ref{Fig_err}(b), as the correlation coefficient $r_{PQ}$ increases, all three error indices exhibit a sharp rise, particularly when $r_{PQ} > 0.9$. 

For \textbf{high collinearity conditions} between $P$ and $Q$ (correlation coefficients of $r\leq0.99$), Fig.~\ref{Fig_8}(e)–(f) show that the Prop.2 method exhibits slightly higher volatility but maintains good unbiasedness. However, under extreme collinearity ($r=0.999$), the method may occasionally yield infeasible results. Both methods significantly outperform existing approaches by reducing estimation errors. As shown in  Fig.~\ref{Fig_err}(c), with increasing delay time $t_{\tau}$, all errors grow steadily. The trends for both methods are similar, but Prop.2 yields consistently smaller errors, with a more pronounced advantage at larger delays.

These findings collectively validate the adaptability and effectiveness of the proposed method in challenging data environments, reinforcing its superiority in practical applications.

\subsection{Practical Engineering Applications}
Unlike MATLAB, the CloudPSS platform \cite{Ref30_song2020cloudpss} 
employs full electromagnetic transient (EMT) simulations, which calculate three-phase electrical quantities in real time and extract current and voltage magnitudes via a phase-locked loop (PLL) to compute power. This process involves multiple signal transformations and numerical computations, which may introduce additional numerical and modeling errors. Moreover, EMT simulations capture more complex electromagnetic phenomena and system dynamics, increasing potential error sources. However, they provide a more accurate representation of the complexity and real behavior of power systems, making them more suitable for certain practical applications. In this section, based on testing on the CloudPSS platform, we further validate the effectiveness of proposed method.

\subsubsection{Case I: Enhanced IEEE 39-Bus System}

In this section, the IEEE 39-bus system \cite{Ref33_athay1979practical} is used as the test case for engineering validation. To reflect the characteristics of modern power systems with high penetration of renewables and dynamic loads, the original IEEE 39-bus system was modified as follows:

\begin{itemize}
    \item \textbf{Photovoltaic Integration:} Two 75\,MW PV units were integrated at Bus~15 and Bus~28, modeled as grid-following inverters with double-loop control (d/q-axis current regulation and outer PQ or voltage-droop control), synchronized via PLL. Output filtering, current limiting, and LVRT protection were included.

    \item \textbf{Wind Power Integration:} Two 100\,MW wind farms were added at Bus~17 and Bus~26, modeled as grid-following full-converter systems with wind-speed-based MPPT and the same inverter control structure as the PV units.
    
    \item \textbf{Dynamic Load Modeling:} The static loads at Bus~4 and Bus~20 were replaced with composite load models (CLMs), each consisting of 50\% induction motor and 50\% ZIP load. The ZIP part used standard coefficients ($A_p = A_q = 0.53$, $B_p = B_q = 0.34$) and the rest as constant impedance.

\end{itemize}

For large-scale EMT systems, it is generally difficult to directly derive the true values of TEP from the detailed network model due to its complexity and opacity. A commonly adopted approach is to inject a step disturbance into the system and use the resulting steady-state responses to estimate reliable reference values. These can serve as benchmarks for validating identification methods based on ambient data.
    
In this study, a sequence of step disturbances with randomly generated magnitudes is superimposed on the active and reactive power at Bus 16. The step signal changes at a frequency of 0.05 Hz, ensuring that voltage, current, and power responses have sufficient time to settle within each interval. This setup produces a set of steady-state operating points under varying load conditions, which are used to robustly estimate high-confidence TEP values.

The estimated reference values (regarded as the “true” TEP for validation) at Bus 16 are $\left| {{E}_{\text{th}}} \right|=500.70\,\text{kV}$, ${{R}_{\text{th}}}=5.62\,\Omega$, and ${{X}_{\text{th}}}=20.45\,\Omega$. 

The load at Bus 16 is modeled as a randomly fluctuating load, as described by \eqref{Eqn28}, with $r_{PQ}$ set to 0.2 and 0.8, respectively. Monte Carlo simulations are conducted, and the resulting TEP identification results for Bus 16, obtained using different methods, are presented in the violin plots in Fig.~\ref{Fig_12}.

\begin{figure}[!htbp] 
    \centering
    \subfloat[]{
		\includegraphics[trim=0.7cm 0.3cm 1.3cm 0.6cm, clip, width=0.48\columnwidth]{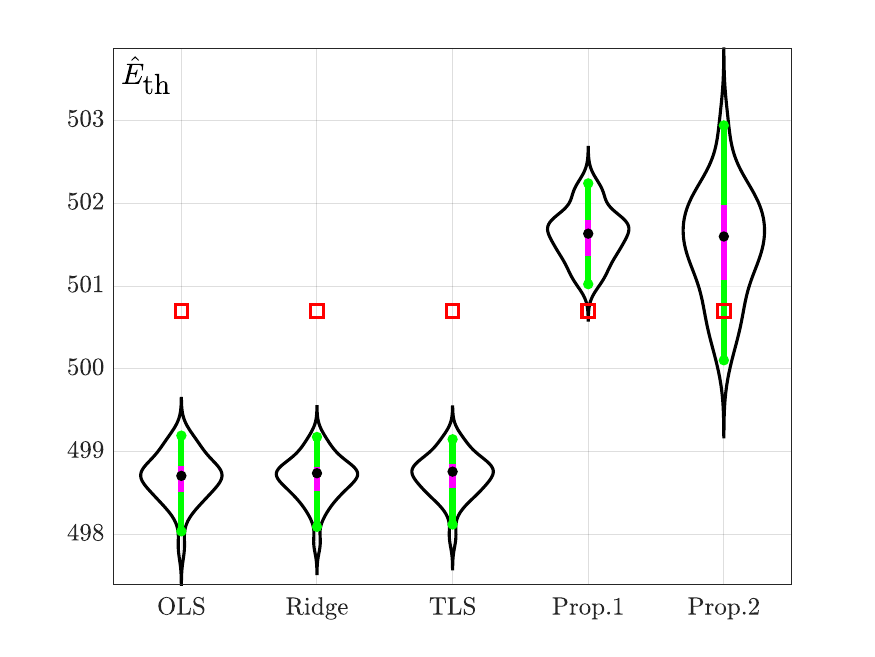}}
	\subfloat[]{
		\includegraphics[trim=1cm 0.3cm 1.3cm 0.6cm, clip, width=0.48\columnwidth]{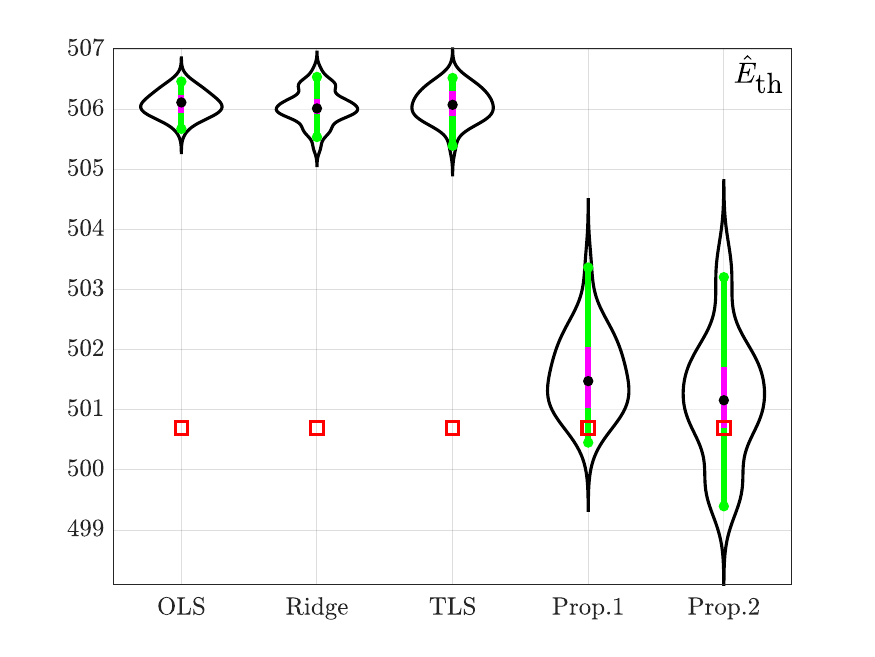}}\\
    \subfloat[]{
		\includegraphics[trim=1cm 0.3cm 1.3cm 0.6cm, clip, width=0.48\columnwidth]{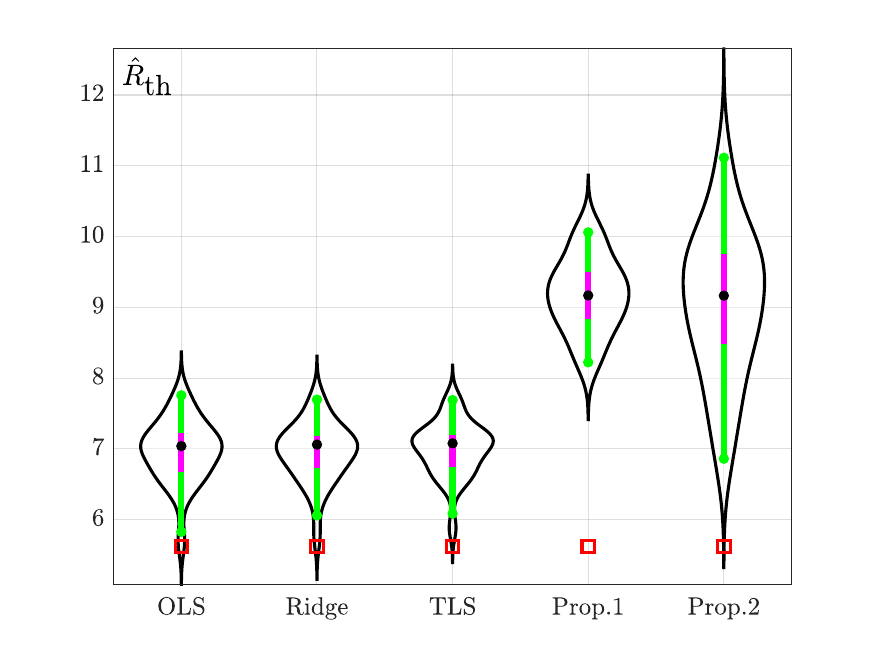}} 
    \subfloat[]{
		\includegraphics[trim=1cm 0.3cm 1.3cm 0.6cm, clip, width=0.48\columnwidth]{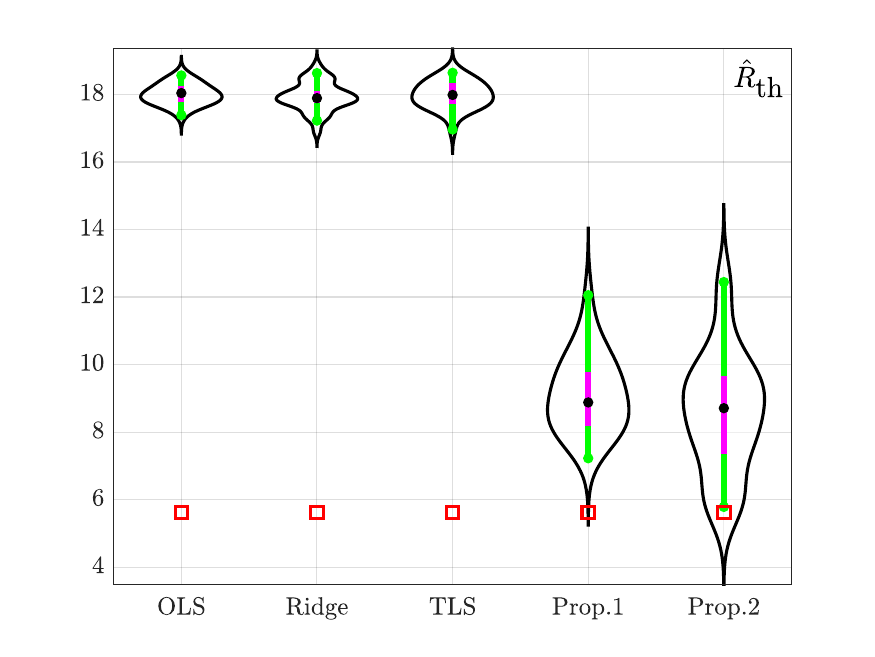}}\\ 
	\subfloat[]{
		\includegraphics[trim=1cm 0.3cm 1.3cm 0.6cm, clip, width=0.48\columnwidth]{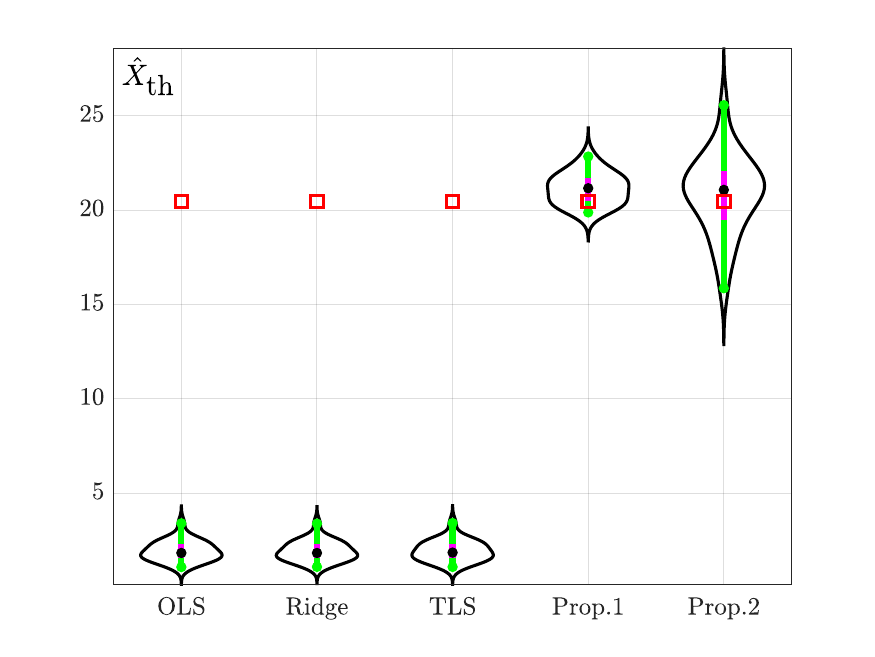}}
    \subfloat[]{
		\includegraphics[trim=1cm 0.3cm 1.3cm 0.6cm, clip, width=0.48\columnwidth]{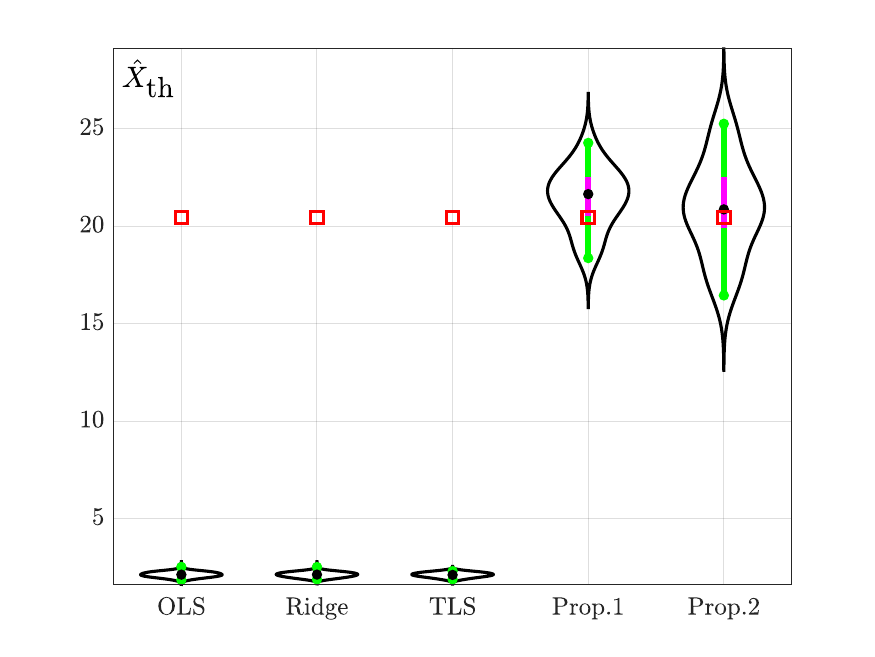}}
    \caption{Violin plots of TEP identification results in enhanced IEEE 39-bus system. Panels (a), (c), and (e) correspond to low $P$–$Q$ correlation ($r_{PQ}=0.2$), while (b), (d), and (f) correspond to high correlation ($r_{PQ}=0.8$). The red box of the violin plots represents the reference values of the parameters. The sampling period $T_{\text{s}}=0.005\,\text{s}$ and the sliding window size $W=5\,\text{s}$. The simulation is performed with 30 independent Monte Carlo experiments, each simulating for 2 minutes. }\label{Fig_12}
\end{figure}
The simulation results in Fig.~\ref{Fig_12} show that, under electromagnetic simulation data conditions, traditional methods perform poorly regardless of whether the load correlation is low or high. TEP identification results from traditional methods exhibit a wide probability distribution with significant estimation bias, making it difficult to accurately match theoretical values.  


Compared to traditional methods, the proposed methods yield probability distributions that are both more concentrated and more centered around the theoretical value of $X_{\mathrm{th}}$, as evidenced by the alignment of the mean and median markers with the ground truth (red square) in Fig.~\ref{Fig_12}. This reflects improved estimation accuracy and robustness.

\subsubsection{Case II: CSEE-RAS System with CLM}

To further verify the robustness of the proposed method under more realistic and large-scale conditions, additional simulations were conducted using the \textbf{CSEE-RAS benchmark system} \cite{Sun2024HybridModel},  designed for rotor angle stability assessment. It is based on a real regional grid in China and comprises 79 buses, multiple AC transmission lines, and one $\pm$500\,kV high-voltage DC link. The CSEE-RAS system features over 50\% renewable energy penetration.

To incorporate dynamic load behavior, the original static load at Bus B02 is replaced with CLM, consisting of 50\% induction motor and 50\% voltage-dependent ZIP components, using the same configuration as described in Case~I. 

The estimated reference values at Bus B01 are $\left| {{E}_{\text{th}}} \right|=229.65\,\text{kV}$, ${{R}_{\text{th}}}=0.78\,\Omega$, and ${{X}_{\text{th}}}=11.40\,\Omega$. Monte Carlo simulations are conducted, and the resulting TEP identification results for Bus B01, obtained using different methods, are presented in the violin plots in Fig.~\ref{Fig_13}.

\begin{figure}[!htbp] 
    \centering
    \subfloat[]{
		\includegraphics[trim=1cm 0.3cm 1.3cm 0.6cm, clip, width=0.48\columnwidth]{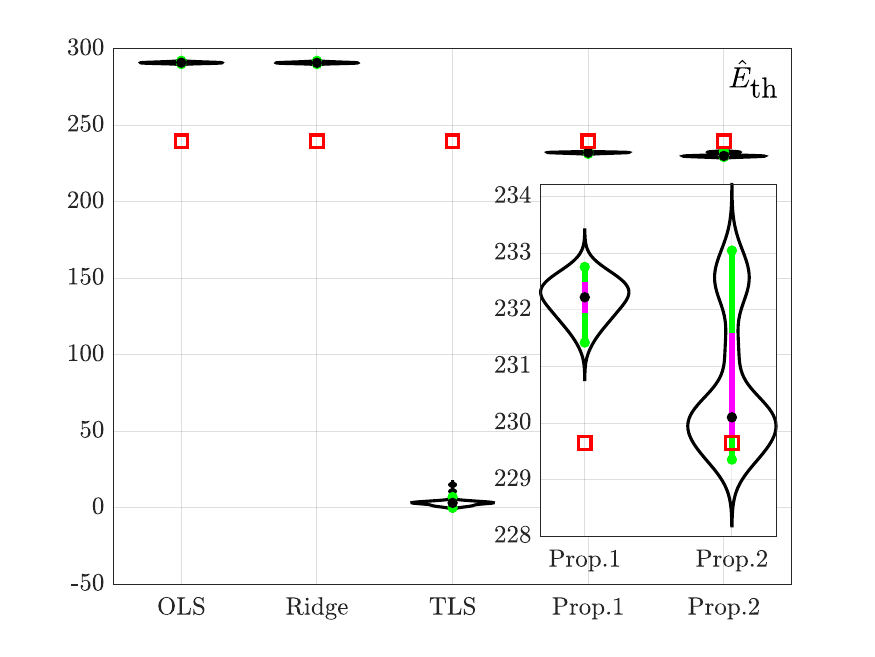}}\\
    \subfloat[]{
		\includegraphics[trim=1cm 0.3cm 1.3cm 0.6cm, clip, width=0.48\columnwidth]{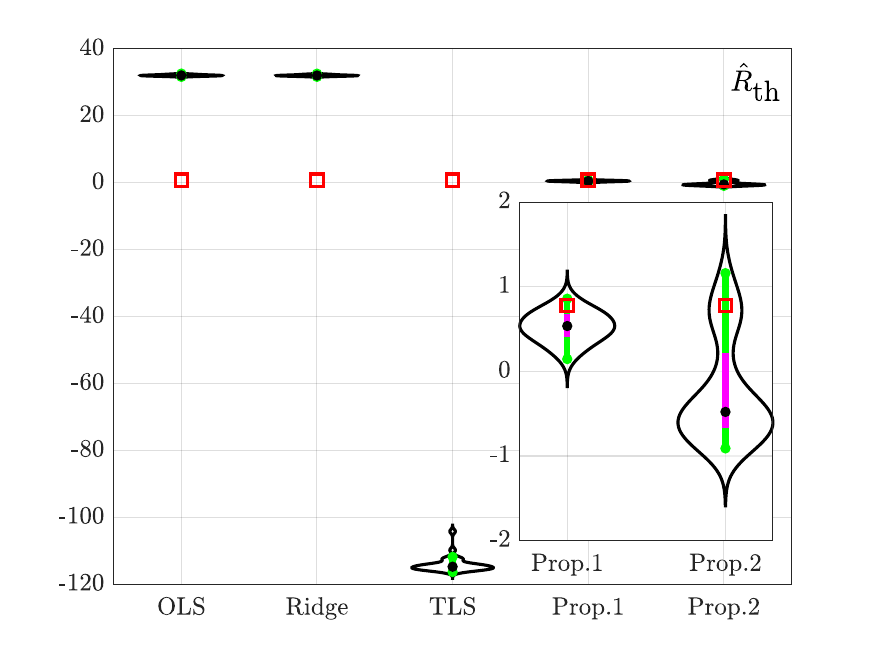}}
	\subfloat[]{
		\includegraphics[trim=1cm 0.3cm 1.3cm 0.6cm, clip, width=0.48\columnwidth]{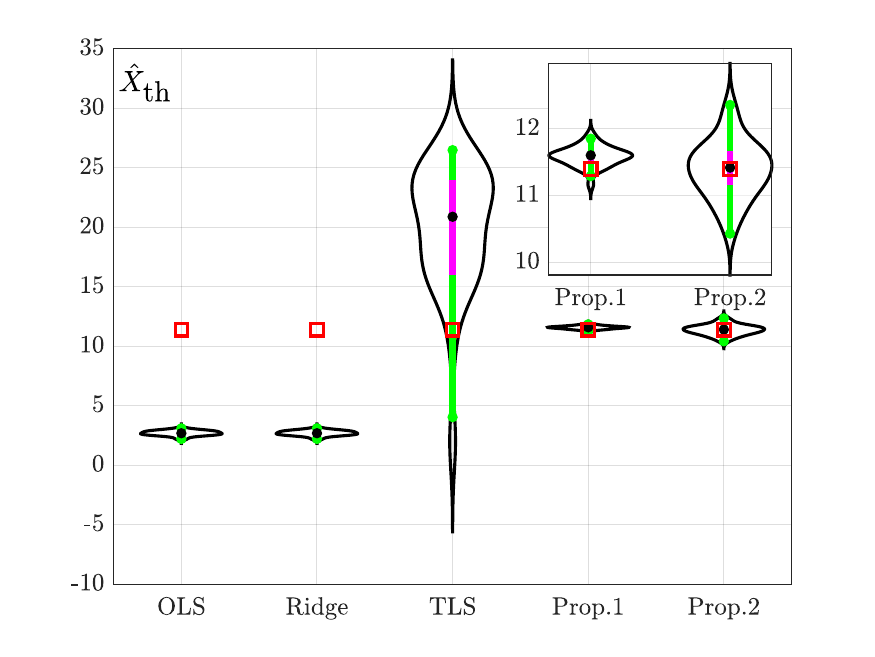}}

    \caption{Violin plots of TEP identification results in the CSEE-RAS system with CLM. $r_{PQ}=0.2$. All simulation settings are consistent with Case~I.}\label{Fig_13}
\end{figure}

The simulation results in Fig.~\ref{Fig_13} similarly validate the robustness of the proposed method under the dynamic stability scenario of the CSEE-RAS system. Despite the increased system complexity and high renewable penetration, the proposed approach maintains accurate and stable TEP estimation performance.

\section{Discussion}

\subsection{Robustness to Non-Gaussian Noise}
The proposed method is primarily evaluated under Gaussian measurement noise, but real-world PMU data may exhibit long-tailed behavior, better modeled by non-Gaussian distributions such as Laplace, Logistic, or Student-$t$ \cite{Wang2018}. These heavy-tailed distributions are more prone to producing outliers or extreme values, which can increase the volatility of statistical characteristics within the sliding window. Consequently, this may theoretically reduce the SNR improvement offered by the proposed method, since the variance of statistical characteristics becomes more sensitive to outliers.

To evaluate the proposed method under these realistic noise conditions, we conduct numerical experiments using the same setup as in Fig.~\ref{Fig_7}(e) of Section~IV. The original Gaussian noise is replaced by different non-Gaussian noise, each adjusted to match the same variance. The results of TEP estimation under these non-Gaussian noise scenarios are shown in Fig.~\ref{Fig_add1}, respectively.

\begin{figure}[!htbp] 
    \centering
    \subfloat[]{
		\includegraphics[trim=0.7cm 0.1cm 1.3cm 0.7cm, clip, width=0.48\columnwidth]{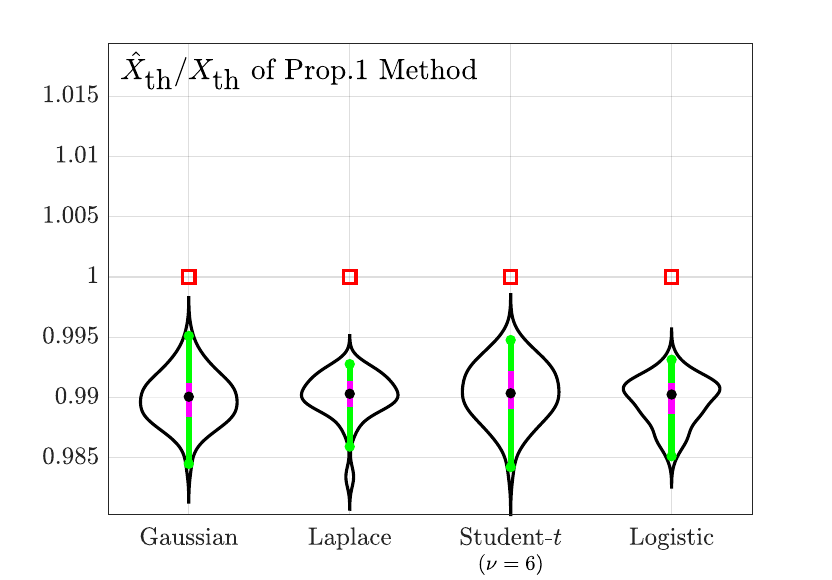}}
        \subfloat[]{
		\includegraphics[trim=0.7cm 0.1cm 1.3cm 0.7cm, clip, width=0.48\columnwidth]{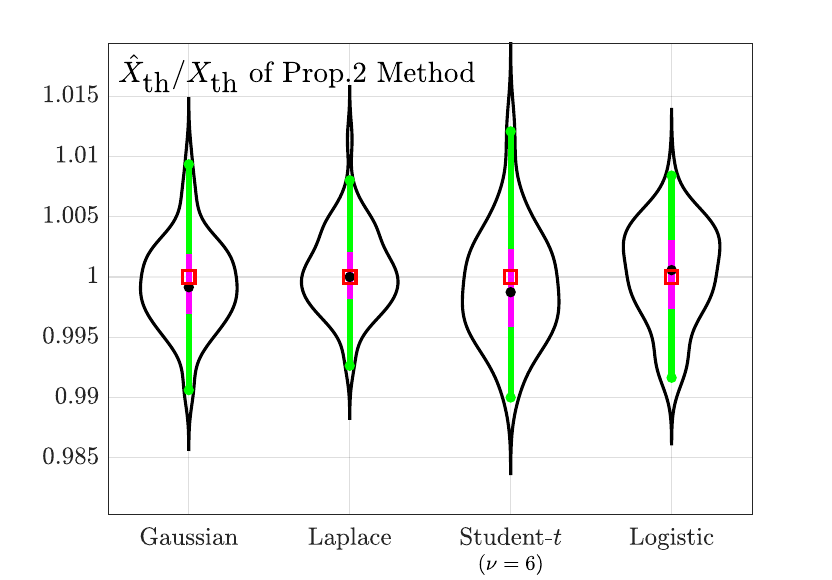}}\\
	\subfloat[]{
		\includegraphics[trim=0.7cm 0.1cm 1.3cm 0.6cm, clip, width=0.48\columnwidth]{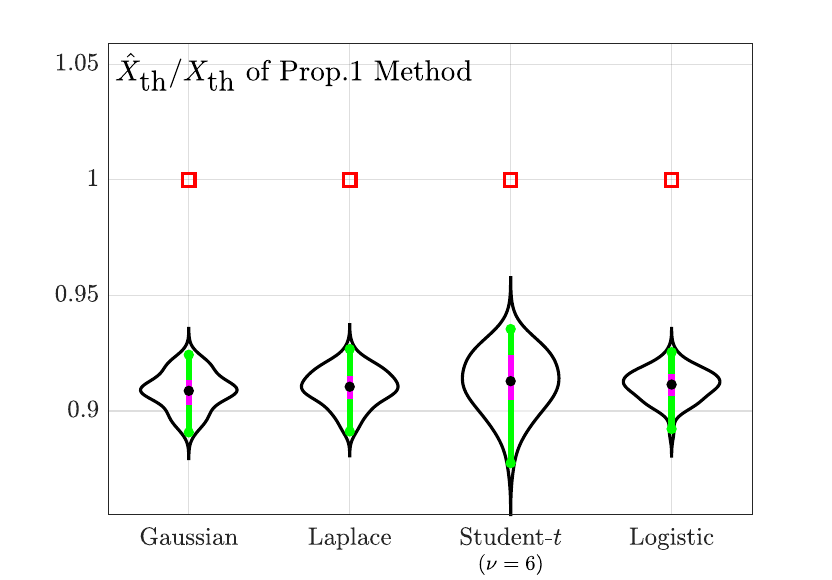}}
    \subfloat[]{
		\includegraphics[trim=0.7cm 0.1cm 1.3cm 0.6cm, clip, width=0.48\columnwidth]{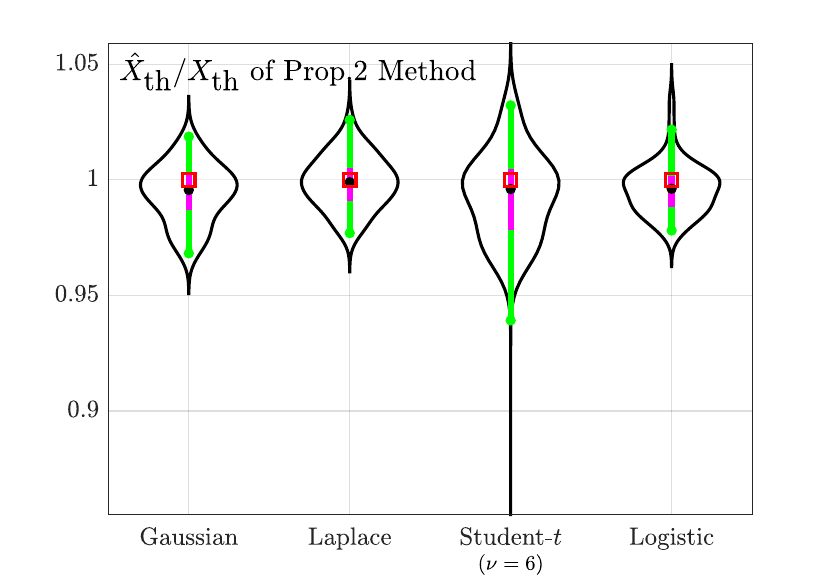}}\\ 
    \caption{Violin plots of $X_{\mathrm{th}}$ identification results using proposed method under different noise models. The load type is a CPL. (a) Prop.1 method: SNR= 20 dB. (b) Prop.2 method: SNR= 20 dB. (c) Prop.1 method: SNR= 10 dB. (d) Prop.2 method: SNR= 10 dB.}\label{Fig_add1}
\end{figure}

As shown in the Fig.~\ref{Fig_add1}, the distribution of the identified parameters under non-Gaussian noise is nearly identical to that under Gaussian noise, demonstrating that the proposed method retains its robustness. This performance stems from the core advantage of the method: exploiting the temporal decorrelation between ambient power fluctuations and measurement noise, as described in \eqref{Eqn18}. This mechanism enables statistical separation of signal and noise components, even when the noise distribution deviates from ideal Gaussian behavior.

\subsection{Robustness to Network Topology Changes}
In practical power systems, $E_{\mathrm{th}}, R_{\mathrm{th}}, X_{\mathrm{th}}$ are not strictly constant due to ongoing network topology changes, even under ambient operating conditions.  In the simulation case, we consider three types of TEP variations: gradual drift, abrupt jumps, and persistent small fluctuations.

To examine the method's performance under these variation patterns, a synthetic 2-hour case study is constructed.  Specifically, $X_{\mathrm{th}}$ increases gradually from 50~$\Omega$ to 80~$\Omega$, simulating a growing electrical distance.  At $t = 3600$~s, $E_{\mathrm{th}}$ undergoes an abrupt jump from 270~kV to 290~kV, representing a power source replacement.  All parameters are further perturbed with zero-mean Gaussian noise to reflect small random fluctuations.
\begin{figure}[!htbp] 
    \centering
    \subfloat[]{
		\includegraphics[trim=0.7cm 0.1cm 0.5cm 0.4cm, clip, width=0.48\columnwidth]{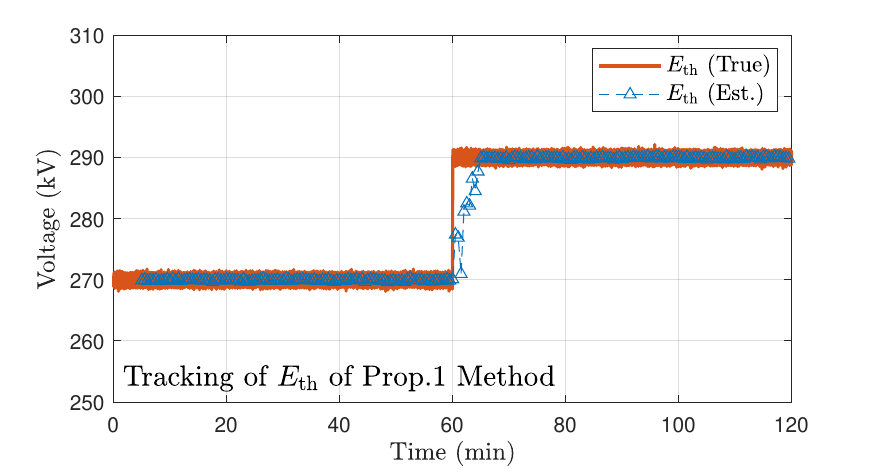}}
        \subfloat[]{
		\includegraphics[trim=0.7cm 0.1cm 0.5cm 0.4cm, clip, width=0.48\columnwidth]{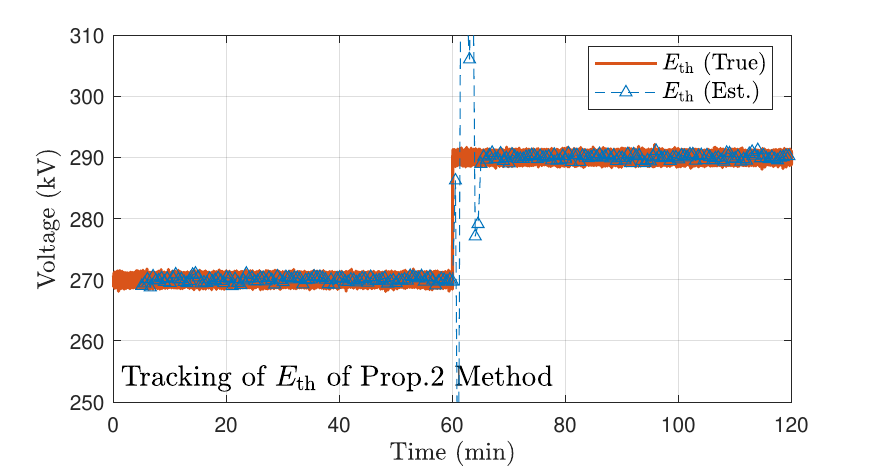}}\\
        \subfloat[]{
    		\includegraphics[trim=0.7cm 0.1cm 0.5cm 0.4cm, clip, width=0.48\columnwidth]{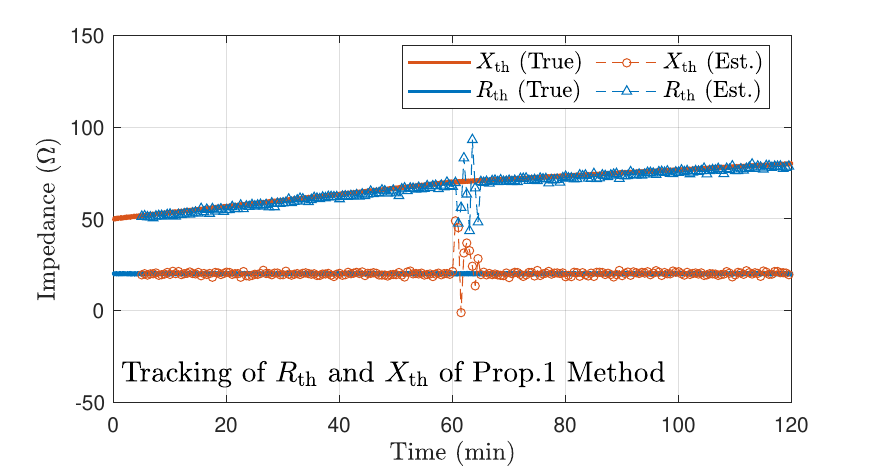}}
        \subfloat[]{
    		\includegraphics[trim=0.7cm 0.1cm 0.5cm 0.4cm, clip, width=0.48\columnwidth]{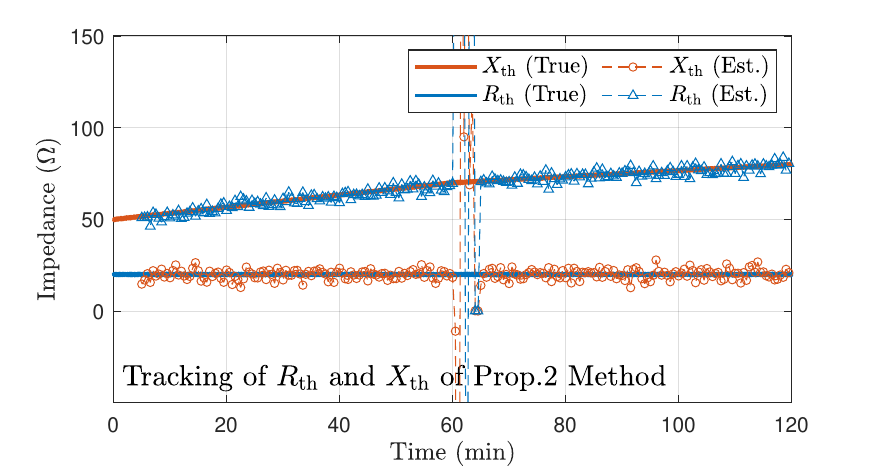}}\\ 
        \caption{Tracking performance of TEP under topology changes using the proposed method with a fixed 5-minute sliding data window.}
    \label{Fig_add4}
    \end{figure}

Fig.~\ref{Fig_add4} presents the tracking results. Both proposed methods recover accurate estimates after abrupt changes. However, Prop.1 demonstrates superior robustness, maintaining smoother and more stable trajectories throughout the process. Prop.2 avoids extreme deviations but tends to fluctuate more under nonstationary conditions. For small fluctuations and gradual drifts, the method captures the local average behavior. For abrupt transitions, the identification may become temporarily inaccurate but recovers promptly as post-event data are incorporated.

Future extensions include incorporating forgetting mechanisms, adaptive estimation, and explicit modeling of topology changes to improve robustness in renewable-rich and actively managed grids.

\subsection{Robustness to Bad Data and Missing Data}
In practical power system operation, PMU measurements often suffer from data imperfections caused by communication delays, sensor faults, cyber attacks, or device aging. To assess the robustness of the proposed identification method under such conditions, we performed a series of tests covering representative forms of bad and missing data.

Asynchronous PMU sampling, a common issue addressed in Sections~III and IV, was complemented by four additional scenarios. as Fig.~\ref{Fig_baddata}:

\begin{figure}[!htbp] 
     \centering
     \subfloat{
         \includegraphics[trim=0.9cm 0.4cm 0.5cm 0.6cm, clip, width=1.0\columnwidth]{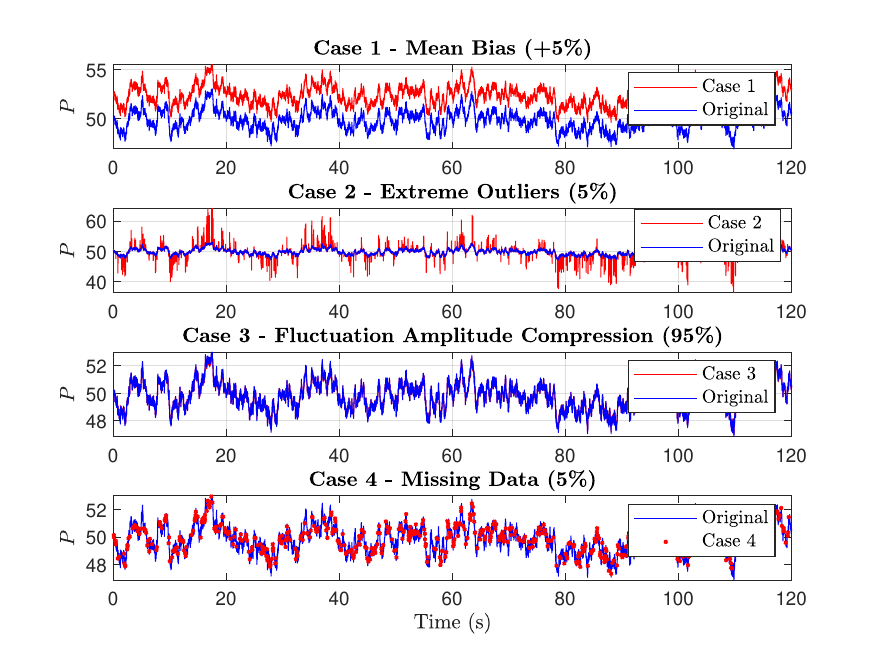}}
     \caption{Bad data and missing data simulation.}\label{Fig_baddata}
 \end{figure}

\subsubsection{Case 1: Sustained Bias in Measurements}
To simulate sensor drift or offset errors, a constant bias of $\pm5\%$ relative to the signal mean was added to $P$, $Q$, $|V|$, and $|I$ measurements. Since MSP features are computed after mean removal, the proposed method remains largely insensitive to this type of error, as shown in Fig.~\ref{Fig_add2}. However, sustained bias in the original signals can still affect the final TEP estimation.

\subsubsection{Case 2: Impulsive Outliers}
To emulate impulsive disturbances, 5\% of the data points were replaced with extreme values defined as $x_i = \mu + 5(x_i - \mu)$. Robust outlier detection was applied using MAD-based thresholds on first-order differences ($x_{i+1} - a x_i$), with outliers replaced by \texttt{NaN} before MSP processing. While this method mitigates most distortions, some residual errors persist, increasing estimation variability, though results generally remain within 10\% error.

\subsubsection{Case 3: Fluctuation Amplitude Compression/Stretching}
This case simulates systematic scaling of signal variance (e.g., due to calibration drift), transforming $P$, $Q$, $|V|$, and $|I|$ as $x_i = \mu + 95\%(x_i - \mu)$ or $105\%(x_i - \mu)$. Though the trend is preserved, the change in amplitude misleads MSP variance estimates and TEP identification. As seen in Fig.~\ref{Fig_add2}, such manipulation significantly biases the results.

\subsubsection{Case 4: Random Missing Data}
To simulate communication loss or transmission dropouts, 5\% of the measurements were randomly replaced with \texttt{NaN}. Sliding-window functions (\texttt{nanmean}, \texttt{nanvar}) inherently ignore \texttt{NaN} values, allowing robust computation of MSP features without interpolation. As shown in Fig.~\ref{Fig_add2}, the impact on TEP estimation is minimal.

\begin{figure}[!htbp] 
    \centering
    \subfloat[]{
        \includegraphics[trim=0.9cm 0.4cm 1.3cm 0.6cm, clip, width=0.48\columnwidth]{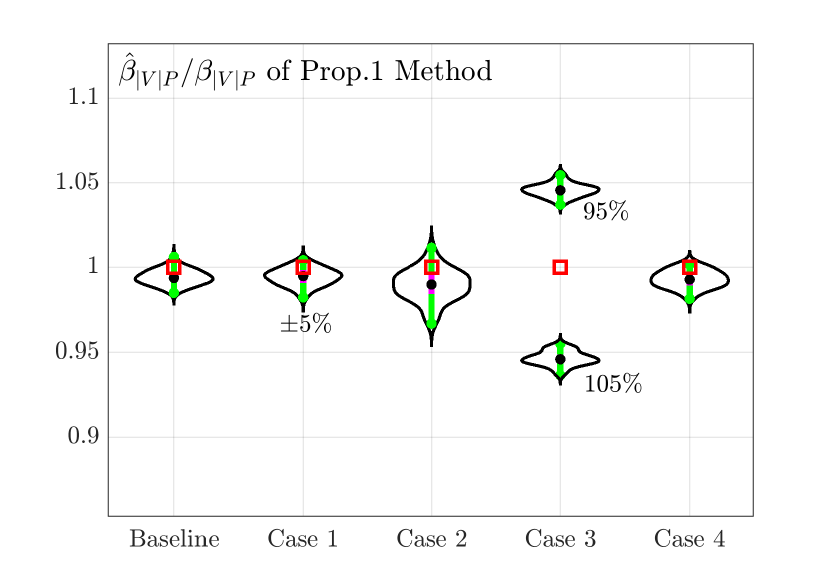}}
        \subfloat[]{
		\includegraphics[trim=0.9cm 0.4cm 1.3cm 0.6cm, clip, width=0.48\columnwidth]{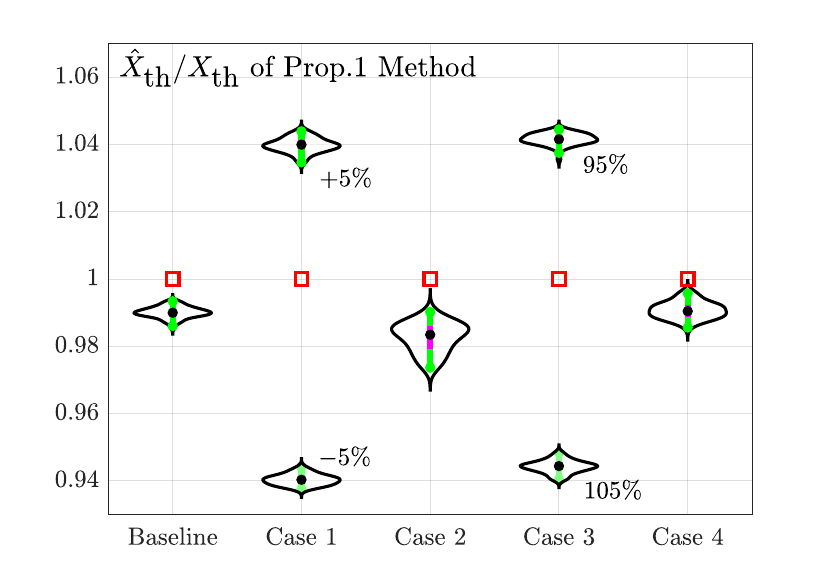}}\\
        \subfloat[]{
		\includegraphics[trim=0.9cm 0.4cm 1.3cm 0.6cm, clip, width=0.48\columnwidth]{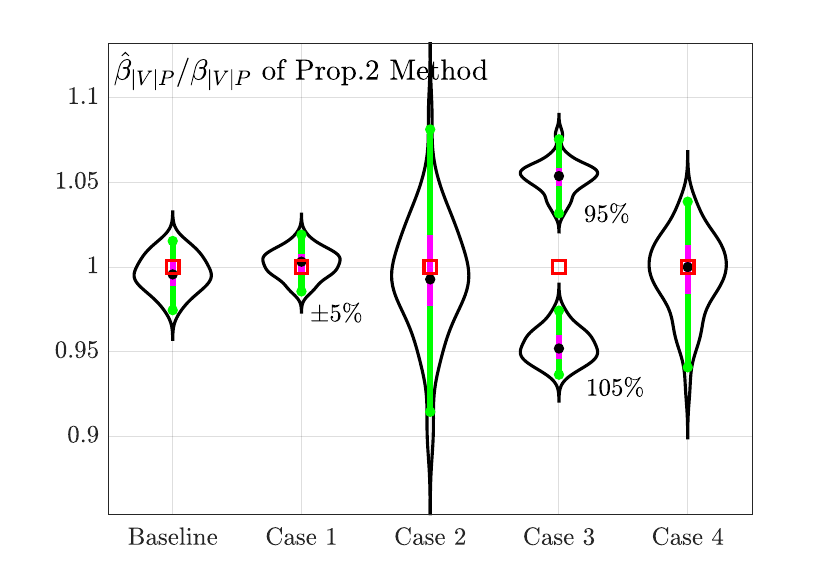}}
        \subfloat[]{
		\includegraphics[trim=0.9cm 0.4cm 1.3cm 0.6cm, clip, width=0.48\columnwidth]{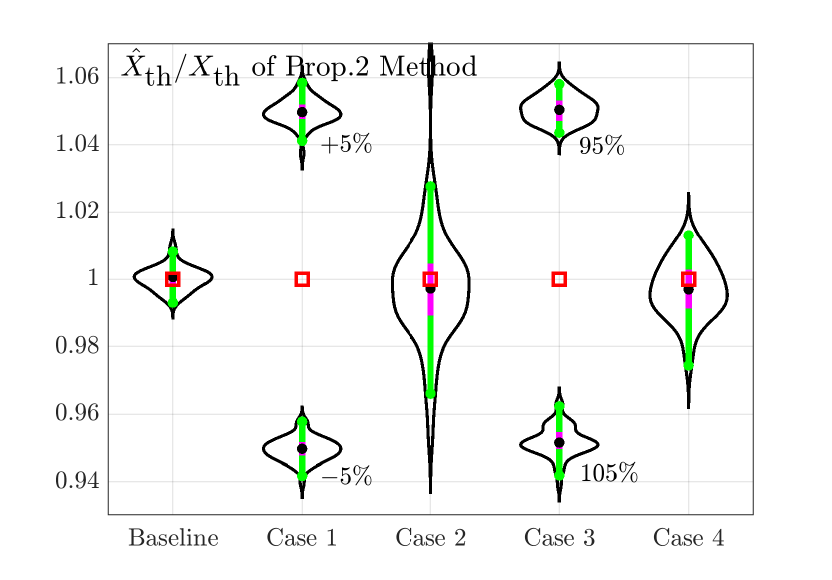}}\\
    \caption{Violin plots of $\beta_{|V|P}$ and $X_{\mathrm{th}}$ identification results using proposed method under different noise models. The Baseline case uses the same configuration as Fig.~\ref{Fig_7}(a).}\label{Fig_add2}
\end{figure}

\begin{figure}[!htbp]
    \centering
    \subfloat[Prop.1 method]{
    \includegraphics[trim=0.5cm 0.7cm 0.3cm 0.6cm, clip, width=0.48\columnwidth]{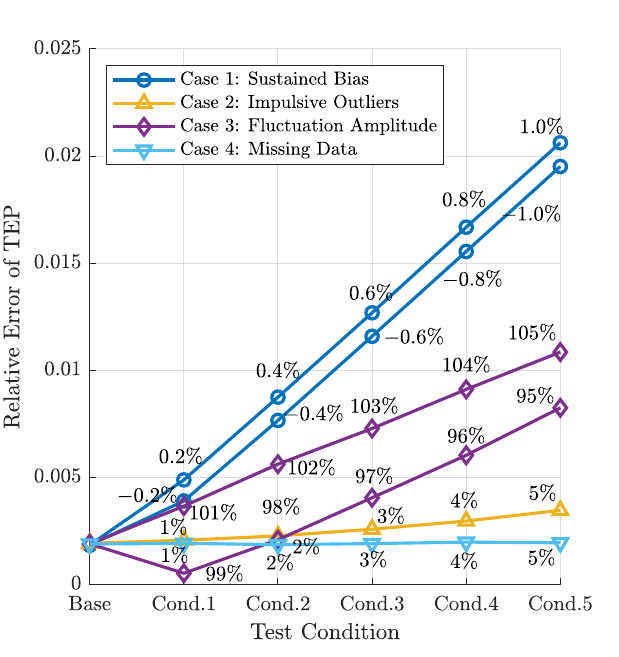}}
    \subfloat[Prop.2 method]{
    \includegraphics[trim=0.5cm 0.7cm 0.3cm 0.6cm, clip, width=0.48\columnwidth]{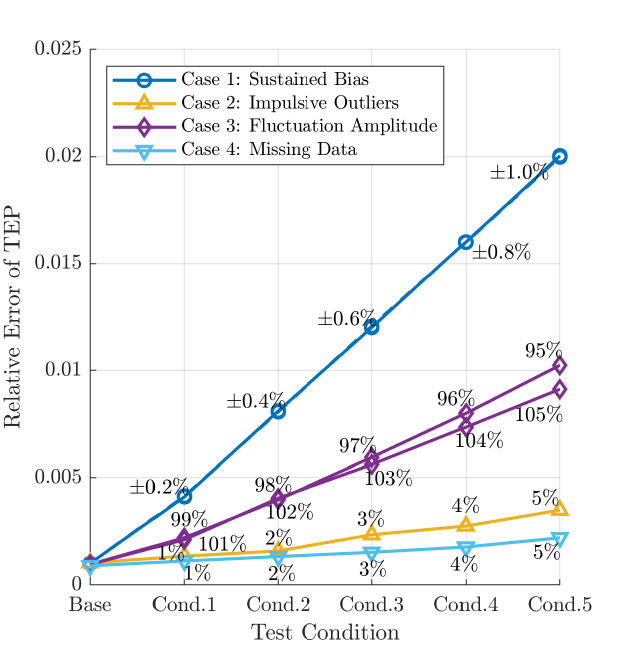}}
    \caption{Relative error of TEP identification under four data corruption cases: (1) Sustained Bias, (2) Impulsive Outliers, (3) Fluctuation Amplitude, and (4) Missing Data. Each curve reflects relative error under increasing severity, computed as $\|\widehat{\boldsymbol{\Theta}} - \boldsymbol{\Theta}_{\mathrm{true}}\|_{\mathrm{F}} / \|\boldsymbol{\Theta}_{\mathrm{true}}\|_{\mathrm{F}}$, averaged over 50 Monte Carlo simulations.}
    \label{Fig_BadData_Err}
\end{figure}

As shown in Fig.~\ref{Fig_BadData_Err}, the proposed method demonstrates strong robustness to randomly occurring anomalies (Cases 2 and 4), while distributional shifts such as mean bias or variance change (Cases 1 and 3) have a greater impact.

Interestingly, in Case~3, the Prop.1 method's error first increases then decreases with stronger compression. This non-monotonic behavior is due to baseline underestimation of $X_{\mathrm{th}}$ being partially offset when fluctuations are further reduced.

Overall, these results confirm the practical applicability of the proposed method in the presence of typical PMU data imperfections. More complex or adversarial scenarios remain as directions for future work.

\section{Conclusion}
This paper presents a novel method for identifying TEP based on the statistical characteristics of system's stochastic response. The method leverages stochastic fluctuation data under steady-state conditions, combined with sliding window techniques, to accurately calculate MSP between voltage magnitude–power and current magnitude–power, thereby achieving high precision and robustness in TEP identification. By analyzing statistical characteristics, such as mean, variance, and covariance, of the stochastic fluctuations within the sliding window, the method enables precise estimation of TEP.

Both theoretical analysis and numerical simulations confirm the effectiveness of the proposed method under challenging scenarios, including low SNR, asynchronous measurements, and high data collinearity in the data. In addition, we discuss the method's robustness under more practical conditions, including non-Gaussian measurement noise, network topology changes, and the presence of bad or missing data. Simulation results further demonstrate that the method maintains strong adaptability and robustness, consistently achieving high identification accuracy despite practical uncertainties encountered in real-world power systems.
 
\bibliographystyle{IEEEtran}

\end{document}